\documentclass[10pt,journal,compsoc]{IEEEtran}
\usepackage[flushleft]{threeparttable}
\usepackage{multirow}
\usepackage{rotating}
\usepackage{booktabs}
\usepackage{amssymb}
\usepackage{amsmath}
\usepackage{amsfonts}
\usepackage{xspace}
\usepackage{graphicx}
\usepackage{mathtools,soul}
\usepackage{subcaption}
\usepackage{enumitem}
\usepackage{bm}
\usepackage{comment}
\usepackage{hyperref}

\newcommand{\etal}{\textit{et al}. }

\newcommand{\method}{\mbox{$\mathop{\mathtt{P^2MAM}}\limits$}\xspace}
\newcommand{\msep}{\mbox{$\mathop{\mathtt{P^2MAM\text{-}P}}\limits$}\xspace}
\newcommand{\mjoint}{\mbox{$\mathop{\mathtt{P^2MAM\text{-}O\text{-}P}}\limits$}\xspace}
\newcommand{\guess}{\mbox{$\mathop{\mathtt{P^2MAM\text{-}O}}\limits$}\xspace}

\newcommand{\ljoint}{\mbox{$\mathop{\mathtt{LAST\text{-}O\text{-}P}}\limits$}\xspace}

\newcommand{\STAMP}{\mbox{$\mathop{\mathtt{STAMP}}\limits$}\xspace}
\newcommand{\LESSR}{\mbox{$\mathop{\mathtt{LESSR}}\limits$}\xspace}
\newcommand{\DHCN}{\mbox{$\mathop{\mathtt{DHCN}}\limits$}\xspace}
\newcommand{\SRGNN}{\mbox{$\mathop{\mathtt{SR\text{-}GNN}}\limits$}\xspace}
\newcommand{\NARM}{\mbox{$\mathop{\mathtt{NARM}}\limits$}\xspace}

\newcommand{\GRURECP}{\mbox{$\mathop{\mathtt{GRU4Rec+}}\limits$}\xspace}
\newcommand{\FGNN}{\mbox{$\mathop{\mathtt{FGNN}}\limits$}\xspace}
\newcommand{\POP}{\mbox{$\mathop{\mathtt{POP}}\limits$}\xspace}

\newcommand{\MEAN}{\mbox{$\mathop{\mathtt{MEAN}}\limits$}\xspace}
\newcommand{\ORACLE}{\mbox{$\mathop{\mathtt{ORACLE}}\limits$}\xspace}

\newcommand{\DG}{\mbox{$\mathop{\texttt{DG}}\limits$}\xspace}
\newcommand{\YC}{\mbox{$\mathop{\texttt{YC}}\limits$}\xspace}
\newcommand{\GA}{\mbox{$\mathop{\texttt{GA}}\limits$}\xspace}
\newcommand{\LF}{\mbox{$\mathop{\texttt{LF}}\limits$}\xspace}
\newcommand{\NP}{\mbox{$\mathop{\texttt{NP}}\limits$}\xspace}
\newcommand{\TM}{\mbox{$\mathop{\texttt{TM}}\limits$}\xspace}

\newcommand{\sr}{\mbox{$\mathop{\mathtt{SR}}\limits$}\xspace}
\newcommand{\ppp}{\mbox{$\mathop{\mathtt{PS}}\limits$}\xspace}
\newcommand{\spp}{\mbox{$\mathop{\mathtt{P^2E}}\limits$}\xspace}
\newcommand{\pe}{\mbox{$\mathop{\mathtt{PE}}\limits$}\xspace}

\newcommand{\nn}{n}

\newcommand{\nh}{b}
\newcommand{\ts}{A}

\newcommand{\pref}{\mbox{$\mathop{\mathbf{h}^o}\limits$}\xspace}
\newcommand{\nei}{\mbox{$\mathop{\mathbf{h}^p}\limits$}\xspace}

\newcommand{\Q}{\mathbf{q}\xspace}
\newcommand{\s}{\mathbf{c}\xspace}

\newcommand{\emb}{\mbox{$\mathop{E}\limits$}\xspace}
\newcommand{\pos}{\mbox{$\mathop{P}\limits$}\xspace}
\newcommand{\V}{\mbox{$\mathop{V}\limits$}\xspace}
\newcommand{\B}{\mbox{$\mathop{C}\limits$}\xspace}

\newcommand{\xia}[1]{\textcolor{red}{#1}}

\ifCLASSOPTIONcompsoc
  \usepackage[nocompress]{cite}
\else
  \usepackage{cite}
\fi
\ifCLASSINFOpdf
\else
\fi
\begin{document}
%
\title{Prospective Preference Enhanced Mixed Attentive Model for Session-based Recommendation}
%
%
%
%

\author{Bo~Peng,~\IEEEmembership{Member,~IEEE,}
        Chang-Yu Tai,
        Srinivasan~Parthasarathy,~\IEEEmembership{Member,~IEEE,}
        and~Xia~Ning$^*$,~\IEEEmembership{Member,~IEEE}
\IEEEcompsocitemizethanks{
\IEEEcompsocthanksitem Bo Peng and Chang-Yu Tai
are with the Department
of Computer Science and Engineering, The Ohio State University, Columbus,
OH, 43210.\protect\\
E-mail: peng.707@buckeyemail.osu.edu, tai.97@buckeyemail.osu.edu 
\IEEEcompsocthanksitem Srinivasan Parthasarathy and Xia Ning are with the Department of Biomedical Informatics, 
and the Department of Computer Science and Engineering, 
The Ohio State University, Columbus, OH, 43210.\protect\\
E-mail:  srini@cse.ohio-state.edu, ning.104@osu.edu
\IEEEcompsocthanksitem $^*$Corresponding author
}
\thanks{Manuscript received April 19, 2005; revised August 26, 2015.}}

%
%

\markboth{Journal of \LaTeX\ Class Files,~Vol.~14, No.~8, August~2015}%
{Shell \MakeLowercase{\textit{et al.}}: Bare Demo of IEEEtran.cls for Computer Society Journals}
%



\IEEEtitleabstractindextext{%
\begin{abstract}
Session-based recommendation aims to generate recommendations for the next item of users' interest based on a given session.
In this manuscript, we develop prospective preference enhanced mixed attentive model (\method) to generate session-based recommendations using two important factors: temporal patterns and estimates of users' prospective preferences.
Unlike existing methods, \method models the temporal patterns using a 
light-weight while effective position-sensitive attention mechanism.
In \method, we also leverage the estimate of users' prospective 
preferences to signify important items, and generate better recommendations.
Our experimental results demonstrate that \method models significantly outperform the state-of-the-art methods in six benchmark datasets, with an improvement as much as 19.2\%.
In addition, our run-time performance comparison demonstrates that during testing, \method models 
are much more efficient than the best baseline method, with a significant average speedup of 47.7 folds.
\end{abstract}

}

\maketitle

\IEEEdisplaynontitleabstractindextext

%
\IEEEpeerreviewmaketitle

\IEEEraisesectionheading{\section{Introduction}\label{sec:intro}}

%
%
%
%
\IEEEPARstart{S}{ession-based} recommendation aims to generate recommendations for the next item of users' interest based 
on a given session (i.e., a sequence of items chronologically ordered according to user interactions in a short-time period).
%
It has been drawing increasing attention from the research community due to its wide applications in 
online shopping~\cite{li2017neural,wu2019session}, music streaming~\cite{xia2020self} and tourist planing~\cite{chen2020handling}, among others.
With the prosperity of deep learning, many deep models, particularly based on recurrent neural networks (RNNs)~\cite{sherstinsky2020fundamentals} 
and graph neural networks (GNNs)~\cite{zhou2020graph}
have been developed for session-based recommendation, and have demonstrated the state-of-the-art performance.
These methods primarily model the temporal patterns (e.g., transitions, recency patterns, etc.) in sessions, 
but are not always effective in modeling other important factors that are indicative of the next item.
In addition, existing methods primarily model the temporal patterns using gated recurrent units (GRUs)~\cite{chung2014empirical}.
However, considering the notorious sparse nature of session-based recommendation datasets, 
as shown in the literature~\cite{peng2020m2}, the complicated GRUs may not be 
well-learned and could degrade the performance.
Due to its recurrent essence, GRUs also suffer from limited parallelizability and poor interpretability.
To mitigate the limitations of existing methods, in this manuscript, we develop prospective preference enhanced mixed attentive model, 
denoted as \method, for session-based recommendation.
%
%
In \method, different from existing methods, we model the temporal patterns using a novel position-sensitive attention mechanism, 
which is light-weight, fully parallelizable, and could enable better interpretability over GRUs. 
Besides the temporal patterns, we also leverage the estimate of users' prospective preferences
 for better recommendation.
Users' prospective preferences is another important factor for recommendation.
Intuitively, if we would have known beforehand that the user is going to
watch action movies next (i.e., prospective preference), 
we could generate better recommendations by learning her/his preference from action movies 
instead of comedy movies in her/his watching history.
We conducted an analysis to empirically verify that users' prospective preferences could signify important items, 
and thus, improve the recommendation performance as in the Discussion Section. 
%
%
The results reveal that conditioned on the prospective preferences, 
we could learn indicative attention weights over items, and enable superior performance.
%
However, in practice, users' prospective preferences are usually intractable.
Thus, in \method, we explicitly estimate the prospective preferences, 
and learn attention weights based on the estimate to boost the recommendation performance.
%

With different combinations of the two factors (i.e., temporal patterns and the estimate of users' prospective preferences),
\method has three variants: \guess, \msep and \mjoint.
\guess models temporal patterns using a novel position-sensitive attention mechanism.
\msep leverages the estimate of users' prospective preferences to weigh items and generate recommendations.
%
\mjoint explicitly leverages the two factors for better recommendation.
%
%

%
%
%
%

%

We compare \method with five state-of-the-art baseline methods on six benchmark session-based recommendation datasets. 
Our experimental results demonstrate that \method significantly outperforms the state-of-the-art methods on all the datasets, with an improvement of up to 19.2\%. 
The results also show that on most of the datasets. the two factors are mutually strengthened, and could enable superior performance when used together.
%
We also conduct a comprehensive analysis to verify the effectiveness of different components in \method. 
The results show that with the position embeddings, our position-sensitive attention mechanism could effectively 
capture the temporal patterns in the datasets, 
and on most of the datasets, our learning-based prospective preference estimate strategy could be more effective than recency-based strategies.
%
Moreover, we conduct run-time performance analysis, 
and find that \method is much more efficient than the best baseline method with an average speedup of 47.7 folds over the six datasets. 
%

Our major contributions are summarized as follows:
\begin{itemize}[leftmargin=*]
\item We develop a novel session-based recommendation method \method, 
which leverages both the temporal patterns and estimates of users' prospective preferences for recommendation.
\item \method significantly outperforms five state-of-the-art methods on six benchmark datasets (Section~\ref{sec:exp:overall}).
\item Our analysis demonstrates the importance of modeling the position information for session-based recommendation (Section~\ref{sec:exp:abla}).
\item The experimental results show that our learning-based prospective preference estimate strategy is more effective than 
the existing recency-based strategy (Section~\ref{sec:exp:global}).
\item Our analysis shows that the learned attention weights in \method could capture the temporal patterns in the data (Section~\ref{sec:exp:att}).
\item Our analysis verifies that users' prospective preferences could signify important items, and benefit recommendations (Section~\ref{sec:dis}).
\item For reproducibility purposes, we release our source code on 
GitHub~\footnote{\url{https://github.com/ninglab/P2MAM}},
and report the hyper parameters in the Appendix.
\end{itemize}

\section{Related Work}
\label{sec:literature}


\subsection{Session-based Recommendation}
\label{sec:literature:session}

%
In the last few years, numerous session-based recommendation methods have been developed, particularly using Markov Chains (MCs), attention mechanisms and neural networks 
such as RNNs and GNNs, etc.
MCs-based methods~\cite{fpmc} use MCs to capture the transitions among items for recommendation.
For example, Rendle \etal~\cite{fpmc} employs a first-order MC to generate recommendations based 
on the transitions of the last item in each session.
Attention-based methods~\cite{li2017neural,liu2018stamp} model the 
importance of items for the recommendation.
%
For example, Liu \etal~\cite{liu2018stamp} developed a short-term attention priority model (\STAMP), 
which adapts a gate mechanism to capture users' short-term preferences.
Recently, RNNs-based methods such as \GRURECP~\cite{hidasi2018recurrent} and \NARM~\cite{li2017neural} have been developed to
model 
the temporal patterns among items primarily using GRUs.
%
%
%

GNNs-based methods are also extensively developed for the session-based recommendation.
Wu \etal~\cite{wu2019session} converted sessions to direct graphs, 
and developed a GNNs-based model (\SRGNN) to generate recommendations based on the graph structures.
%
Qiu \etal~\cite{qiu2019rethinking} re-examined the importance of item ordering in session-based recommendations 
and developed a GNN-based model (\FGNN), which included self-loop for each node in graphs to better capture users' short-term preferences.
%
Chen \etal~\cite{chen2020handling} showed that the widely used directed graph representations
can not fully preserve the sequential information in sessions.
To mitigate this problem, they converted sessions to multigraphs, and developed a GNNs-based model (\LESSR) to generate recommendations.
%
Xia \etal~\cite{xia2020self} developed hyper graph-based model (\DHCN), which leverages hyper graphs and hyper graph convolutional networks to 
capture the high-order information among items.
%

%

\subsection{Sequential Recommendation}
\label{sec:literature:sequential}

Sequential recommendation aims to generate recommendations for the next items of users' interest based on users’ historical interactions.
It is closely related to session-based recommendation except that in sequential recommendation, 
we could access users' historical interactions in a long-time period (e.g., months).
%
In the last few years, neural networks 
(e.g., RNNs) and attention mechanisms have been extensively employed in sequential recommendation methods.
For example, RNNs-based methods such as User-based RNNs~\cite{ugru} incorporate user characteristics into GRUs for personalized recommendation. 
Skip-gram-based methods~\cite{vasile2016meta}
leverage the skip-gram model~\cite{mikolov2013distributed} to capture the co-occurrence among items in a time window.
Recently, Convolutional Neural Networks (CNNs) are also adapted for sequential recommendation.
Tang \etal~\cite{tang2018personalized} developed a CNNs-based model, which uses multiple convolutional filters to model the synergies~\cite{peng2021ham} among items. 
Yuan \etal~\cite{yuan2019simple} developed another CNN-based generative model NextItRec to better capture the long-term dependencies in sequential recommendation.
Besides CNNs, attention-based methods~\cite{kang2018self,sun2019bert4rec,fan2021modeling} are also developed for sequential recommendation.
Kang \etal~\cite{kang2018self} developed a self-attention based model, which adapts the self-attention to better model users' long-term preferences. 
Sun \etal~\cite{sun2019bert4rec} further developed a bidirectional self-attention based model
to improve the representational power of item embeddings.
%

\section{Definitions and Notations}
\label{sec:define}

\begin{table}[!t]
  \caption{Notations}
  \label{tbl:notations}
  \centering
  \begin{threeparttable}
     \begin{footnotesize}
      \begin{tabular}{
	@{\hspace{16pt}}l@{\hspace{16pt}}
	@{\hspace{16pt}}l@{\hspace{16pt}}          
	}
        \toprule
        notations & meanings \\
        \midrule
        $m$    &  the number of items \\
        $d$     &  the dimensionality of  embeddings\\
        $S$     & the original session\\
        $\ts$    & transformed fixed-length sequence\\
        $\nn$   & the number of items in $\ts$\\
        $\pref$ & position-sensitive preference prediction\\
        $\nei$ & prospective preferences-sensitive prediction\\        
        $\hat{\mathbf{r}}$ & recommendation scores\\
        \bottomrule
      \end{tabular}
      \end{footnotesize}
  \end{threeparttable}
\end{table}


In this manuscript, we tackle the recommendation problem 
that given an anonymous session,
we recommend the next item of users' interest in the session. 
%
An anonymous session
 is represented as a sequence 
$S = \{s_1, s_2, \dots\, s_{|S|}\}$, where $s_t$ is the $t$-th item in the session and $|S|$ is the length of the session.
We use upper-case letters to denote matrices, lower-case and bold letters to denote
row vectors, and lower-case non-bold letters to denote scalars. 
Table~\ref{tbl:notations} presents the key notations used in this manuscript.

\section{Methods}
\label{sec:method}

\begin{figure}[t]
    \includegraphics[width=0.75\linewidth,angle=270]{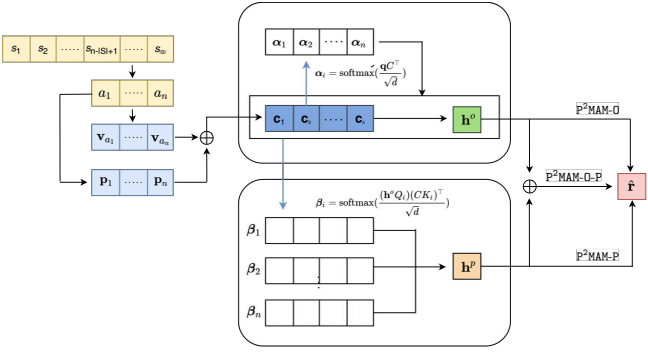}
    \vspace{-20pt}
    \caption{Overall Architecture}
    \label{fig:architecture}
\end{figure}

Figure~\ref{fig:architecture} presents the overall architecture of \method. 
\method generates recommendations via a session representation component, a position-sensitive preference prediction component, 
and a prospective preference enhanced preference prediction component.
We will discuss each component in detail below. 

\subsection{Session Representation (\sr)}
\label{sec:method:embed}

Previous studies~\cite{kang2018self,peng2021ham,tang2018personalized} have shown that 
recently interacted items are more indicative than earlier ones of the next item.
%
Following these,
we focus on the most recent $\nn$ items (i.e., the last $\nn$ items) in a session to generate recommendations.
Particularly, given a session $S = \{s_1, s_2, \dots\, s_{|S|}\}$, 
we transform it to a fixed-length sequence $\ts = \{a_{1}, a_{2}, \dots, a_{\nn} \}$, 
which contains the last $\nn$ items in $S$ (i.e., $a_i$ = $s_{|S|-n+i}$, $i = 1, \cdots, n$). 
%
If $S$ is shorter than \nn, we will pad empty items at the beginning of $\ts$ until length \nn.

In \method, we represent the items in sessions using learnable embeddings.
Specifically, we learn an item embedding matrix $\V \in \mathbb{R}^{m \times d}$, 
in which the $j$-th row $\mathbf{v}_j$ is the embedding of item $j$, and $m$ and $d$ is 
the number of all the items and the dimensionality of embeddings, respectively.
Given $\V$, we represent the items in $\ts$
by a matrix $\emb \in \mathbb{R}^{\nn \times d}$ as follows: 
\begin{equation}
    \label{eqn:emb}
    \emb = [\mathbf{v}_{a_{1}}; \mathbf{v}_{a_{2}}; \cdots; \mathbf{v}_{a_{\nn}}],
\end{equation}
where $\mathbf{v}_{a_{i}}$ is the embedding of item $a_{i}$. 
Following the previous work~\cite{vaswani2017attention,kang2018self,fan2021modeling}, 
we use a constant zero vector as the embedding of padded empty items. 

We also learn embeddings for the $\nn$ positions in the fixed-length sessions to capture the temporal patterns.
Specifically, following Vaswani \etal~\cite{vaswani2017attention}, we learn a position embedding matrix $\pos \in \mathbb{R}^{\nn \times d}$:
\begin{equation}
    \label{eqn:pos}
    \pos = [\mathbf{p}_{1}; \mathbf{p}_{2}; \cdots; \mathbf{p}_{\nn}],
\end{equation}
in which the $t$-th row $\mathbf{p}_{t}$ is the embedding of the $t$-th position. 


\subsection{\mbox{Position-sensitive Preference Prediction (\ppp)}}
\label{sec:method:predict}

%

It has been shown~\cite{peng2020m2,liu2018stamp,peng2021ham} that the temporal patterns play an important role
in predicting users' preferences.
%
Existing session-based recommendation methods model the temporal patterns primarily using GRUs-based methods~\cite{chung2014empirical,li2015gated}, 
which implicitly learn weights over items.
However, as demonstrated in the literature~\cite{peng2020m2}, the complicated GRUs-based models may not be well learned
for the notoriously sparse recommendation datasets, 
and it also suffers from poor interpretability~\cite{hou2020learning} and limited parallelizability~\cite{peng2020m2,kang2018self}.

%

In \method, different from other methods, 
we develop a novel position-sensitive attention mechanism to 
model the temporal patterns, 
and generate predictions of users' preferences accordingly.
%
Specifically, we use a dot-product attention mechanism as follows:
\vspace{-3pt}
\begin{eqnarray}
    \label{eqn:dot}
    \begin{aligned}
        \B = \emb + \pos, && 
        \boldsymbol{\alpha}  = \text{softmax} ( \frac{\Q \B^\top}{\sqrt{d}} ), && 
        \pref &=  \boldsymbol{\alpha} \B,
     \end{aligned}
     \vspace{-5pt}
\end{eqnarray}
%
where 
$\B$ combines items embeddings and position embeddings in the session,
and $\s_i$ is the $i$-th row in $\B$, 
$\boldsymbol{\alpha}$ is a vector of attention weights,
$\Q \in \mathbb{R}^{1 \times d}$ is a learnable vector shared by all the sessions, and
\pref is the position-sensitive preference prediction.
%
The intuition of our position-sensitive preference prediction is that 
items themselves (their contents) and their relative orders in the sessions represent meaningful 
information related to user preferences; we can use data-driven attention weights 
from item embeddings and position embeddings to capture such information.
%
%
%
Different from existing attentive session-based methods~\cite{liu2018stamp,li2017neural,xu2019graph}, 
which learn attention weights without explicitly considering the position information,
%
we explicitly incorporate position embeddings to learn position-sensitive attention weights.
%
%
Compared to GRUs-based methods~\cite{chung2014empirical,li2015gated}, 
the simple and light-weight attention mechanism is easier to learn well on the sparse recommendation datasets, 
and it could also provide better interpretability and parallelizability.

%

%

\subsection{Prospective Preference Enhanced Preference Prediction (\spp)}
\label{sec:method:items}

As presented in Section~\ref{sec:intro}, 
users' prospective preferences reveal important items, 
and thus, could benefit the recommendation.
%
Motivated by this, in \method, 
we develop a novel strategy that leverages the preference prediction (i.e., \pref) from \ppp as 
an estimate of users' prospective preferences 
to enable better attention weights over items.
Specifically, we employ a multi-head attention mechanism~\cite{vaswani2017attention} as follows:
%
\begin{eqnarray}
    \label{eqn:multi}
    \begin{aligned}
       \boldsymbol{\beta}_i &= \text{softmax}(\frac{(\pref Q_i) (\B K_i)^\top}{\sqrt{d}}), \\
        \mathbf{head}_i &= \boldsymbol{\beta}_i (\B W_i), \\
        \nei &= [\mathbf{head}_1, \mathbf{head}_2, \dots, \mathbf{head}_{\nh}] W,\\
    \end{aligned}
    \vspace{-10pt}
\end{eqnarray}
where $\boldsymbol{\beta}_i$ is the vector of attention weights from the $i$-th head, 
$Q_i \in \mathbb{R}^{d \times \frac{d}{\nh}}$, $K_i \in \mathbb{R}^{d \times \frac{d}{\nh}}$, 
and $W_i \in \mathbb{R}^{d \times \frac{d}{\nh}}$ are learnable projection matrices in the $i$-th head,
and $\mathbf{head}_i$ is the output of the $i$-th head, 
$\nh$ is the number of heads,
$W \in \mathbb{R}^{d \times d}$ is a learnable projection matrix shared by all the heads, 
and $\nei \in \mathbb{R}^{1 \times d}$ is the prospective preference enhanced preference prediction. 
%
The intuition of our strategy (i.e., \spp) is that we learn predictive attention weights based on the 
estimate of users' prospective preferences. 
As will be shown in Section~\ref{sec:exp:overall}, 
this strategy could significantly improve the recommendation performance.
Note that \spp serves as a general strategy that is adaptable to other estimates of 
users' prospective preferences.
%
%
We also tried the dot-product attention as in \ppp but empirically found that it produces inferior performance.

%


\subsection{Recommendation Scores in \method}
\label{sec:method:rec}

In \method, we calculate recommendation scores based on the predictions of users' preferences (i.e., $\pref$ and $\nei$).
Specifically, we develop three different methods to calculate the scores.
Based on the scores, the items with the top-$k$ largest scores will be recommended.

\subsubsection{Scores based on temporal patterns (\guess)}
\label{sec:method:rec:guess}

%
Similarly to existing methods~\cite{li2017neural,wu2019session}, 
we calculate the recommendation scores based on the temporal patterns as follows:
\begin{equation}
    \label{eqn:score:guess}
    \hat{\mathbf{r}} = \text{softmax} (\pref \V^\top), 
\end{equation}
where 
$\hat{\mathbf{r}}$ is a vector of recommendation scores over candidate items, 
$\V$ is the item embedding matrix (Section~\ref{sec:method:embed}), 
and the softmax function is employed to normalize the scores to be into the range $[0, 1]$.
%
We denote this \method variant as \guess.
%

\subsubsection{Scores based on users' prospective preferences (\msep)}
\label{sec:method:rec:sep}

In \method, we could also generate recommendations from 
the prospective preference enhanced preference prediction $\nei$ as follows:
\begin{equation}
    \label{eqn:score:sep}
    \hat{\mathbf{r}} = \text{softmax} (\nei \V^\top). 
\end{equation}
%
We denote this \method variant as \msep.

\subsubsection{Scores based on temporal patterns 
and users' prospective preferences (\mjoint)}
\label{sec:method:rec:joint}

The two preference predictions (i.e., $\pref$ and $\nei$) could be mutually strengthened
and might enable better performance when used together. 
%
Following this motivation, we also calculate recommendation scores using both $\pref$ and $\nei$ as follows:
\begin{equation}
    \label{eqn:score:joint}
    \hat{\mathbf{r}}= \text{softmax} ( (\pref + \nei ) \V^\top),
\end{equation}
where we sum $\pref$ and $\nei$ for the recommendation.
We denote the \method variant using both $\pref$ and $\nei$ as \mjoint.
%


\subsection{Network Training}
\label{sec:method:train}

Following the literature~\cite{liu2018stamp,wu2019session,chen2020handling}, 
we adapt the cross-entropy loss
to minimize the negative log likelihood of correctly recommending the ground-truth next item as follows:
\begin{equation}
    \label{eqn:obj:primary}
    \min\limits_{\boldsymbol{\Theta}} \sum \nolimits_{i=1}^{|T|} -\mathbf{r}_i \log(\hat{\mathbf{r}}_i^\top), 
\end{equation}
where $T$ is the set of all the training sessions, $\mathbf{r}_i$ is a one-hot vector in which the dimension $j$ is 1 if item $j$ 
is the ground-truth next item in the $i$-th training session or 0 otherwise, 
$\hat{\mathbf{r}}_i$ is the vector of recommendation scores for the $i$-th training session,
and $\Theta$ is the set of learnable parameters (e.g., $\V$, $\pos$ and $W$). 
All the learnable parameters are randomly initialized, 
and are optimized in an end-to-end manner.
%
We use this objective to optimize all the \method variants (i.e., \guess, \msep and \mjoint).

\section{Materials}
\label{sec:material}

\subsection{Baseline Methods}
\label{sec:material:baseline}

We compare \method with the five state-of-the-art baseline methods:
\begin{itemize}[leftmargin=*]
\item \POP~\cite{wu2019session} recommends the most popular items of each session.
\item \NARM~\cite{li2017neural} employs GRUs and attention mechanisms to model the temporal patterns for the recommendation.
\item \SRGNN~\cite{wu2019session} transforms sessions into directed graphs, and employs GNNs to 
model complex transitions in sessions.
\item \LESSR~\cite{chen2020handling} transforms sessions into directed multigraphs 
and generates recommendations using GNNs. 
\item \DHCN~\cite{xia2020self} transforms sessions into hypergraphs
and generates recommendations 
using a hypergraph convolutional network.
\end{itemize}
%
%
Note that \DHCN and \LESSR have been compared against a comprehensive set of other methods 
including \GRURECP~\cite{hidasi2018recurrent}, \STAMP~\cite{liu2018stamp} and \FGNN~\cite{qiu2019rethinking}, 
and have outperformed those methods.
Thus, we compare \method with \DHCN and \LESSR instead of the methods that they have outperformed. 
For all the baseline methods, we use the implementations provided by their authors (Section~\ref{app:rep} in Appendix).

\subsection{Datasets}
\label{sec:material:datasets}

We compare \method with the baseline methods in the following benchmark datasets that are 
widely used in the literature~\cite{li2017neural,wu2019session,chen2020handling}:
\begin{itemize}[leftmargin=*]
\item Diginetica (\DG)~\footnote{\url{ http://cikm2016.cs.iupui.edu/cikm-cup}} is from the CIKM Cup 2016, and contains anonymized browsing logs and transactions.
Following the literature~\cite{li2017neural,wu2019session,chen2020handling}, we only use the transaction data in our experiments.
\item Yoochoose (\YC)~\footnote{\url{http://2015.recsyschallenge.com/challenge.html}} is from the RecSys Challenge 2015 containing sessions of user clicks within 6 months.
\item Gowalla (\GA)~\footnote{\url{https://snap.stanford.edu/data/loc-gowalla.html}} is a point-of-interests dataset and includes user-venue check-in records with timestamps.
\item Lastfm (\LF)~\footnote{\url{http://ocelma.net/MusicRecommendationDataset/lastfm-1K.html}} is a dataset collecting streams of music listening events in the Last.fm.
Following the literature~\cite{chen2020handling}, we focus on the music artist recommendation in our experiments.
\item Nowplaying (\NP)~\footnote{\url{ http://dbis-nowplaying.uibk.ac.at/\#nowplaying}} is another dataset describing the music listening events of users.
\item Tmall (\TM)~\footnote{\url{https://tianchi.aliyun.com/dataset/dataDetail?dataId=42}} is from the IJCAI-15 competition, and it includes 
anonymized shopping logs on the online retail platform Tmall.
\end{itemize}
Following the literature~\cite{li2017neural,wu2019session,chen2020handling,xia2020self}, for \GA, 
we only keep the top 30,000 most popular locations, and view the check-in records in one day as a session~\cite{chen2020handling,peng2020m2}.
For \LF, we only keep the top 40,000 most popular artists, and view the listening events in 8 hours as a session~\cite{chen2020handling}.
For all the datasets, we filter out sessions of length one and items appearing less than five times over all the sessions.
%

\subsection{Experimental Protocol}
\label{sec:material:protocol}


\begin{table}[!t]
  \caption{Dataset Statistics}
  \label{tbl:data}
  \centering
  \begin{threeparttable}
      \begin{footnotesize}
      \begin{tabular}{
	@{\hspace{1pt}}l@{\hspace{1pt}}
        @{\hspace{1pt}}r@{\hspace{1pt}}
        @{\hspace{2pt}}r@{\hspace{2pt}}
        @{\hspace{2pt}}r@{\hspace{2pt}}
        @{\hspace{2pt}}r@{\hspace{2pt}}
        @{\hspace{1pt}}r@{\hspace{1pt}}
        @{\hspace{1pt}}r@{\hspace{1pt}}
        @{\hspace{2pt}}r@{\hspace{2pt}}
        @{\hspace{2pt}}r@{\hspace{2pt}}
        @{\hspace{2pt}}r@{\hspace{0pt}}
	}
        \toprule
        dataset & \#items && \#train & \#test & length   && \#aug train & \#aug test & aug len\\
        \midrule
        DG     & 42,596  && 188,636 & 15,955 & 4.80   && 716,835    & 60,194  & 4.90\\
        YC     & 17,597  && 124,472 & 15,237 & 4.22    && 394,802   & 55,424  & 6.14\\
        GA     & 29,510  && 234,403 & 57,492 & 3.85    && 675,561   & 155,332& 4.32\\
        LF      & 38,615  && 260,780 & 64,763 & 11.78  && 2,837,644 & 672,519& 9.16\\
        NP     & 60,416  && 128,077 & 14,479 & 7.42    && 825,304    & 89,824 & 6.53\\
        TM     & 40,727  &&   65,286 &  1,027  & 6.69    && 351,268    & 25,898  & 8.01\\
        \bottomrule
      \end{tabular}
      \end{footnotesize}
      \begin{tablenotes}
        \setlength\labelsep{0pt}
	\begin{footnotesize}
	\item
        In this table, \#item is the number of items.
        The columns \#train, \#test, and length correspond to the number of training sessions, 
        the number of testing sessions, and the average length of sessions, respectively, before the augmentation.
        The columns \#aug train,  \#aug test and `aug len' correspond to that after the augmentation.
        \par
	\end{footnotesize}
      \end{tablenotes}
  \end{threeparttable}
\end{table}


\subsubsection{Training and testing Sets}
\label{sec:material:protocol:train}

Following the literature~\cite{wu2019session,xia2020self,chen2020handling}, we generate the training and testing sets as follows:
for \DG, \NP and \TM, we use the sessions in the last week as the testing set, and all the other sessions as the training set.
For \YC, we use the sessions in the last day as the testing set.
For the other sessions, following that in Li \etal~\cite{li2017neural}, we use the last (i.e., most recent) $1/64$ of them for training. 
For \GA and \LF, we use the last (i.e., most recent) 20\% of all the sessions for testing, and all the other sessions for training.
%

Following the literature~\cite{li2017neural,wu2019session,xia2020self,chen2020handling}, 
we augment the data to enrich the training and testing data.
%
Specifically, 
for each original training and testing session $S=\{s_1, s_2, \dots, s_{|S|}\}$, 
we split it to $\{s_1, s_2\}$, $\{s_1, s_2, s_3\}$, $\cdots$, $\{s_1, s_2, \dots, s_{|S|-1}\}$ and $\{s_1, \dots, s_{|S|}\}$, 
and use all the resulted sessions as the augmented sessions for training and testing.
%
%
The key statistics of the original and augmented datasets are presented in Table~\ref{tbl:data}.
Note that, we transform sessions to fixed-length as in Section~\ref{sec:method:embed} after the augmentation.

We tune the hyper parameters using grid search and use the best hyper parameters in terms of recall@20 (Section~\ref{sec:material:protocol:metric}) 
for \method and all the baseline methods during testing.
Particularly, during the hyper parameter tuning, we use the first 80\% of training sessions for model training, 
and evaluate the model on the last 20\% of training sessions.
During testing, we use all the training sessions for model training, and evaluate methods in testing sessions.
%
We report the search ranges of hyper parameters, and the identified optimal hyper parameters for each method in the Appendix (Section~\ref{app:rep}).

\subsubsection{Evaluation metrics}
\label{sec:material:protocol:metric}


We use recall@$k$, MRR@$k$ and NDCG@$k$ to evaluate the performance of methods.

\begin{itemize}[leftmargin=*]
\item Recall@$k$ measures the proportion of sessions in which the ground-truth next item (i.e., $s_{|S|+1}$) is correctly recommended.
For each session $S$, the recall@$k$ is 1 if $s_{|S|+1}$ is among the top $k$ of the 
recommendation list, or 0 otherwise. 
Note that in next item recommendation,  
recall@$k$ is the most popular evaluation metric.
It is also called precision@$k$ and HR@$k$ as in the literature~\cite{chen2020handling}.
%
%
\item MRR@$k$ is the mean reciprocal rank of the correctly recommended item, 
and is 0 if the ground-truth next item is not in the top-$k$ of the recommendation list.
MRR@$k$ is widely used in the literature~\cite{li2017neural,wu2019session,xia2020self,chen2020handling,liu2018stamp} 
as a rank-aware evaluation metric for session-based recommendation.
\item NDCG@$k$ is the normalized discounted cumulative gain for the top-k ranking, 
and is another widely used rank-aware metric~\cite{tang2018personalized,fan2021continuous,kang2018self}.
%
Different from MRR@$k$ that only focuses on the very top ranked items (e.g., top-$1$)~\cite{wu2011optimizing},
NDCG@$k$ effectively measures models' performance in ranking the top-$k$ items, and thus,  
might be a better metric for evaluating recommendation methods in some scenarios. 
%
Follow the literature~\cite{tang2018personalized}, In our experiments, the gain indicates whether the ground-truth next item is recommended (i.e., gain is 1) or not (i.e., gain is 0).
%
\end{itemize}
For all the evaluation metrics, we report the average results over all the testing sessions in the experiments. 
%
We also statistically test the significance of the performance difference among different methods via a standard paired $t$-test
at 95\% confidence level.

\section{Experimental Results}
\label{sec:exp}


\subsection{Overall Performance Comparison}
\label{sec:exp:overall}

\begin{table*}[h]
  \caption{Overall Performance} 
  \label{tbl:performance}
  \centering
  \begin{threeparttable}
     \begin{scriptsize}
      \begin{tabular}{
	    @{\hspace{0pt}}l@{\hspace{1pt}}
	    @{\hspace{1pt}}c@{\hspace{1pt}}
	    @{\hspace{2pt}}r@{\hspace{2pt}}
            @{\hspace{2pt}}r@{\hspace{2pt}}
            @{\hspace{1pt}}r@{\hspace{1pt}}
            @{\hspace{2pt}}r@{\hspace{2pt}}
            @{\hspace{2pt}}r@{\hspace{2pt}}
            @{\hspace{1pt}}r@{\hspace{1pt}}
           @{\hspace{2pt}}r@{\hspace{2pt}}
           @{\hspace{2pt}}r@{\hspace{2pt}}
           @{\hspace{1pt}}r@{\hspace{1pt}}
           @{\hspace{1pt}}c@{\hspace{1pt}}
           @{\hspace{2pt}}r@{\hspace{2pt}}
            @{\hspace{2pt}}r@{\hspace{2pt}}
            @{\hspace{1pt}}r@{\hspace{1pt}}
            @{\hspace{2pt}}r@{\hspace{2pt}}
            @{\hspace{2pt}}r@{\hspace{2pt}}
            @{\hspace{1pt}}r@{\hspace{1pt}}
           @{\hspace{2pt}}r@{\hspace{2pt}}
           @{\hspace{2pt}}r@{\hspace{2pt}}
            @{\hspace{1pt}}r@{\hspace{1pt}}
            @{\hspace{1pt}}c@{\hspace{1pt}}
            @{\hspace{2pt}}r@{\hspace{2pt}}
            @{\hspace{2pt}}r@{\hspace{2pt}}
            @{\hspace{1pt}}r@{\hspace{1pt}}
            @{\hspace{2pt}}r@{\hspace{2pt}}
            @{\hspace{2pt}}r@{\hspace{2pt}}
            @{\hspace{1pt}}r@{\hspace{1pt}}
           @{\hspace{2pt}}r@{\hspace{2pt}}
           @{\hspace{2pt}}r@{\hspace{0pt}}
	}
        \toprule
        \multirow{2}{*}{method} & \multirow{2}{*}{} & \multicolumn{2}{c}{recall@$k$} 
        && \multicolumn{2}{c}{MRR@$k$} && \multicolumn{2}{c}{NDCG@$k$} 
        && \multirow{2}{*}{} & \multicolumn{2}{c}{recall@$k$} 
        && \multicolumn{2}{c}{MRR@$k$} && \multicolumn{2}{c}{NDCG@$k$} 
        && \multirow{2}{*}{} & \multicolumn{2}{c}{recall@$k$} 
        && \multicolumn{2}{c}{MRR@$k$} && \multicolumn{2}{c}{NDCG@$k$} \\
        \cline{3-4} \cline{6-7} \cline{9-10} \cline{13-14} \cline{16-17} \cline{19-20} \cline{23-24}
        \cline{26-27} \cline{29-30}
        && $k$=10 & $k$=20 && $k$=10 & $k$=20 && $k$=10 & $k$=20 &&
        & $k$=10 & $k$=20 && $k$=10 & $k$=20 && $k$=10 & $k$=20 &&
        & $k$=10 & $k$=20 && $k$=10 & $k$=20 && $k$=10 & $k$=20\\
        \midrule
        \POP & \multirow{8}{*}{\rotatebox[origin=c]{90}{\DG}}
        & 0.0058 & 0.0078 && 0.0021 & 0.0022 && 0.0029 & 0.0034 
        && \multirow{8}{*}{\rotatebox[origin=c]{90}{\YC}}
        & 0.0555 & 0.1102 && 0.0248 & 0.0285 && 0.0319 & 0.0456
        && \multirow{8}{*}{\rotatebox[origin=c]{90}{\GA}}
        & 0.0266 & 0.0456 && 0.0066 & 0.0079 && 0.0113 & 0.0160\\
        \NARM && 0.4049 & \underline{0.5401} && 0.1751 & 0.1845 && 0.2289 & 0.2631
        &&& 0.5970 & 0.7054 && 0.2925 & 0.3001 && 0.3648 & 0.3924
        &&& 0.4510 & 0.5321 && 0.2445 & 0.2502 && 0.2939 & 0.3144\\
        \SRGNN && 0.3783 & 0.5091 && 0.1629 & 0.1719 && 0.2133 & 0.2463
        &&& 0.5979 & 0.7072 && 0.2969 & 0.3046 && 0.3683 & 0.3961
        &&& 0.4317 & 0.5142 && 0.2389 & 0.2446 && 0.2848 & 0.2057\\
        \LESSR && 0.4000 & 0.5321 && 0.1765 & 0.1857 && 0.2289 & 0.2623
        &&& \underline{0.6098} & 0.7140 && \underline{$\mathclap{^{\dagger~}}$0.3077} & \underline{$\mathclap{^{\dagger~}}$0.3151} 
        && \underline{$\mathclap{^{\dagger~}}$0.3795} & \underline{$\mathclap{^{\dagger~}}$0.4060}
        &&& 0.4440 & 0.5229 && \underline{0.2540} & \underline{0.2595} && \underline{0.2993} & \underline{0.3192}\\
        \DHCN && \underline{0.4058} & 0.5400 && \underline{0.1768} & \underline{0.1861} && \underline{0.2305} & \underline{0.2644}
        &&& 0.6090 & \underline{0.7157} && 0.2964 & 0.3039 && 0.3708 & 0.3979
        &&& \underline{0.4531} & \underline{0.5377} && 0.2354 & 0.2413 && 0.2876 & 0.3089\\
        \guess && \textbf{$\mathclap{^{\dagger~}}$0.4148} & \textbf{$\mathclap{^{\dagger~}}$0.5500} && \textbf{$\mathclap{^{\dagger~}}$0.1817} & \textbf{$\mathclap{^{\dagger~}}$0.1911} 
        && \textbf{$\mathclap{^{\dagger~}}$0.2364} & \textbf{$\mathclap{^{\dagger~}}$0.2705}
        &&& 0.6118 & 0.7203 && 0.2956 & 0.3033 && 0.3707 & 0.3983
        &&& 0.4529 & 0.5352 && 0.2384 & 0.2441 && 0.2898 & 0.3106\\
        \msep && 0.4028 & 0.5354 && 0.1749 & 0.1841 && 0.2283 & 0.2618
        &&& 0.6148 & 0.7220 && \textbf{0.3021} & \textbf{0.3096} && 0.3764 & 0.4037
        &&& 0.4554 & 0.5369 && 0.2493 & 0.2550 && 0.2985 & 0.3192\\
        \mjoint && 0.4120 & 0.5474 && 0.1805 & 0.1899 && 0.2348 & 0.2690
        &&& \textbf{$\mathclap{^{\dagger~}}$0.6192} & \textbf{$\mathclap{^{\dagger~}}$0.7277} && 0.3008 & 0.3085 && \textbf{0.3766} & \textbf{0.4043}
        &&& \textbf{$\mathclap{^{\dagger~}}$0.4644} & \textbf{$\mathclap{^{\dagger~}}$0.5479} && \textbf{$\mathclap{^{\dagger~}}$0.2586} & \textbf{$\mathclap{^{\dagger~}}$0.2644} 
        && \textbf{$\mathclap{^{\dagger~}}$0.3077} & \textbf{$\mathclap{^{\dagger~}}$0.3289}\\
        \cmidrule(lr){3-10} \cmidrule(lr){13-20} \cmidrule(lr){23-30}
        improv && 2.2\%$\mathclap{^*}$ & 1.8\%$\mathclap{^*}$ 
        && 2.8\%$\mathclap{^*}$ & 2.7\%$\mathclap{^*}$ 
        && 2.6\%$\mathclap{^*}$ & 2.3\%$\mathclap{^*}$ 
        &&& 1.5\%$\mathclap{^*}$ & 1.7\%$\mathclap{^*}$ 
        && -1.8\%$\mathclap{^*}$ & -1.7\%$\mathclap{^*}$ 
        && -0.8\% & -0.4\%
        &&& 2.5\%$\mathclap{^*}$ & 1.9\%$\mathclap{^*}$ && 1.8\%$\mathclap{^*}$ & 1.9\%$\mathclap{^*}$ 
        && 2.8\%$\mathclap{^*}$ & 3.0\%$\mathclap{^*}$\\
        \midrule
        \POP & \multirow{8}{*}{\rotatebox[origin=c]{90}{\LF}} 
        & 0.0304 & 0.0498 && 0.0113 & 0.0127 && 0.0157 & 0.0207
        && \multirow{8}{*}{\rotatebox[origin=c]{90}{\NP}}
        & 0.0137 & 0.0171 && 0.0058 & 0.0060 && 0.0075 & 0.0084
        && \multirow{8}{*}{\rotatebox[origin=c]{90}{\TM}}
        & 0.0177 & 0.0231 && 0.0095 & 0.0099 && 0.0114 & 0.0128\\
        \NARM && 0.1632 & 0.2287 && 0.0720 & 0.0765 && 0.0933 & 0.1099
        &&& 0.1494 & 0.2037 && 0.0702 & 0.0739 && 0.0887 & 0.1024
        &&& \underline{0.2489} & 0.2661 && \underline{$\mathclap{^{\dagger~}}$0.1765} & \underline{$\mathclap{^{\dagger~}}$0.1777} 
        && \underline{$\mathclap{^{\dagger~}}$0.1944} & \underline{0.1987}\\
        \SRGNN && 0.1666 & 0.2260 && 0.0853 & 0.0893 && 0.1043 & 0.1192
        &&& 0.1436 & 0.1893 && 0.0744 & 0.0775 && 0.0906 & 0.1021
        &&& 0.2438 & \underline{0.2885} && 0.1366 & 0.1397 && 0.1621 & 0.1734\\
        \LESSR &&  \underline{0.1719} & \underline{0.2328} && \underline{$\mathclap{^{\dagger~}}$0.0865} & \underline{$\mathclap{^{\dagger~}}$0.0907} 
        && \underline{$\mathclap{^{\dagger~}}$0.1065} & \underline{$\mathclap{^{\dagger~}}$0.1218}
        &&& 0.1542 & 0.2059 && 0.0747 & 0.0782 && 0.0933 & 0.1063
        &&& 0.2216 & 0.2567 && 0.1267 & 0.1291 && 0.1493 & 0.1582\\
        \DHCN && 0.1647 & 0.2293 && 0.0730 & 0.0774 && 0.0945 & 0.1107
        &&& \underline{0.1711} & \underline{0.2307} && \underline{$\mathclap{^{\dagger~}}$0.0750} & \underline{$\mathclap{^{\dagger~}}$0.0791} 
        && \underline{$\mathclap{^{\dagger~}}$0.0974} & \underline{0.1124}
        &&& 0.2330 & 0.2839 && 0.1294 & 0.1329 && 0.1539 & 0.1668\\
        \guess && 0.1646 & 0.2301 && 0.0724 & 0.0769 && 0.0940 & 0.1105
        &&& \textbf{$\mathclap{^{\dagger~}}$0.1744} & \textbf{$\mathclap{^{\dagger~}}$0.2371} && 0.0740 & \textbf{0.0783} && \textbf{0.0973} & \textbf{$\mathclap{^{\dagger~}}$0.1132}
        &&& \textbf{$\mathclap{^{\dagger~}}$0.2840} & 0.3406 && \textbf{0.1608} & \textbf{0.1648} 
        && \textbf{0.1900} & \textbf{$\mathclap{^{\dagger~}}$0.2043}\\
        \msep && 0.1675 & 0.2334 && 0.0743 & 0.0788 && 0.0961 & 0.1127
        &&& 0.1587 & 0.2132 && \textbf{0.0743} & 0.0780 && 0.0940 & 0.1078
        &&& 0.2302 & 0.2772 && 0.1224 & 0.1256 && 0.1479 & 0.1598\\
        \mjoint && \textbf{$\mathclap{^{\dagger~}}$0.1771} & \textbf{$\mathclap{^{\dagger~}}$0.2454} && \textbf{0.0793} & \textbf{0.0840} && \textbf{0.1022} & \textbf{0.1195}
        &&& 0.1688 & 0.2315 && 0.0732 & 0.0775 && 0.0955 & 0.1113
        &&& 0.2826 & \textbf{$\mathclap{^{\dagger~}}$0.3438} && 0.1484 & 0.1527 && 0.1801 & 0.1956\\
        \cmidrule(lr){3-10} \cmidrule(lr){13-20} \cmidrule(lr){23-30}
        improv && 3.0\%$\mathclap{^*}$ & 5.4\%$\mathclap{^*}$ 
        && -8.3\%$\mathclap{^*}$ & -7.4\%$\mathclap{^*}$ 
        && -4.0\%$\mathclap{^*}$ & -1.9\%$\mathclap{^*}$
        &&& 1.9\% & 2.8\%$\mathclap{^*}$ && -0.9\% & -1.0\% && -0.1\% & 0.7\%
        &&& 14.1\%$\mathclap{^*}$ & 19.2\%$\mathclap{^*}$ && -8.9\%$\mathclap{^*}$ & -7.3\%$\mathclap{^*}$ && -2.3\% & 2.8\%\\

        \bottomrule
      \end{tabular}
      \end{scriptsize}
      \begin{tablenotes}
        \setlength\labelsep{0pt}
	\begin{footnotesize}
	\item
         For each dataset, the best performance among our proposed methods (e.g., \mjoint) is in \textbf{bold}, 
         the best performance among the baseline methods is \underline{underlined}, 
         and the overall best performance is indicated by a dagger (i.e.,  $\dagger$). 
         The row ”improv” presents the percentage improvement of the best performing variant of \method (\textbf{bold}) 
         over the best performing baseline methods (\underline{underlined}). 
         The $*$ indicates that the improvement is statistically significant at $95\%$ confidence level.
        \par
	\end{footnotesize}
      \end{tablenotes}
  \end{threeparttable}
\end{table*}




Table~\ref{tbl:performance} presents the overall performance of different methods at recall@$k$, MRR@$k$ and NDCG@k in recommending the next item. 
Due to the space limit, we do not present the results on recall@5, MRR@5 and NDCG@5.
However, we observed a similar trend on these metrics.
Table~\ref{tbl:performance} shows that overall, \method (i.e., \guess, \msep and \mjoint) is the best performing method on the six benchmark datasets.
In terms of recall@10 and recall@20, 
\method achieves the best performance on all the six datasets, 
with significant average improvement of 4.2\% and 5.5\%, respectively, compared to the best baseline method at each dataset.
Note that in session-based recommendation, improvement of above 1\% is generally considered as significant~\cite{wu2019session,chen2020handling}.
%
%
At MRR@$k$, \method also achieves competitive performance over the baseline methods.
For example, 
on \DG and \GA, \method achieves statistically significant improvement of 2.7\% and 1.9\% at MRR@10 and MRR@20, respectively.
On \YC and \TM, \method achieves the second best performance at MRR@10 and MRR@20.
We found a similar trend at NDCG@$k$. 
For example, in terms of NDCG@$10$, \method substantially outperforms the baseline methods on \DG and \GA; 
at NDCG@$20$, \method is the best method on four out of six datasets except \YC and \LF.
These results demonstrate the strong recommendation performance of \method.
We notice that on \YC, \LF and \TM, at MRR@10 and MRR@20, \method appears considerably worse than the baseline methods (e.g., \LESSR). 
These results indicate that on certain datasets, 
even though \method may be less effective than the baseline methods 
on ranking the ground-truth next items on the very top (e.g., top-1), 
\method is still on average more effective on recommending the correct items among on top.

%

%
%
%

\subsection{Comparison among \method Variants}
\label{sec:exp:variant}

As shown in Table~\ref{tbl:performance}, among the three variants of \method,
\mjoint has the best performance overall.
In terms of recall@10, \mjoint outperforms the other variants on \YC, \GA and \LF, 
and achieves the second best performance on the other three datasets. 
In terms of recall@20, \mjoint is the best method at four out of six datasets except \DG and \NP.
%
We found a similar trend at MRR@$k$ and NDCG@$k$.
For example, in terms of NDCG@10 and NDCG@20, 
\method achieves significant improvement over
 the other variants on three out of six datasets (i.e., \YC, \GA and \LF). 
On the other three datasets, 
{\mjoint} is still ranked as the second best method.
%

Compared to \guess, \mjoint learns attention weights conditioned on the estimate of users' prospective preferences (i.e., \spp), 
while \guess does not have this strategy.
The superior 
performance of \mjoint over \guess on four out of six datasets indicates that on most of the datasets, 
incorporating the estimate of prospective preferences could enable better recommendations.
%
Compared to \msep, \mjoint generates recommendations using the preference predictions from temporal patterns (i.e., $\pref$) and users' prospective preferences (i.e., $\nei$), 
while \msep only use $\nei$ to generate recommendations.
%
%
The strong improvement of \mjoint compared to \msep indicates that 
when used together, the two preference predictions 
could reinforce each other and improve the recommendation performance.
%
We notice that on \DG and \NP, the performance of \mjoint is slightly worse than that of \guess at recall@20.
This might be due to that as will be shown in Section~\ref{sec:exp:abla}, 
in certain datasets (e.g. \DG and \NP), the temporal patterns are highly strong, 
and could denominate the learning process.
As a result, incorporating the \spp component may not improve the recommendation performance.



%
%

\subsection{Comparison with GRUs-based Methods}
\label{sec:exp:gru}

Existing methods~\cite{li2017neural,wu2019session,chen2020handling} leverage GRUs to capture the temporal patterns, 
while in \method, we model the temporal patterns using a position-sensitive attention mechanism (i.e., \ppp). 
Here, we compare the performance of \guess, the \method variant purely based on \ppp, 
and the GRUs-based baseline methods including \NARM, \SRGNN and \LESSR.
As shown in Table~\ref{tbl:performance}, \guess achieves superior performance over GRUs-based baseline methods 
on five out of six datasets except \LF at recall@$k$.
On \LF, the performance of \guess is slightly worse than that of \SRGNN and \LESSR 
but \guess is still able to outperform \NARM at all the metrics.
%
GRUs model the temporal patterns in a recurrent fashion using complicated non-linear layers, 
while 
\guess explicitly models the temporal patterns using attention weights over items in the session.
%
The significant improvement of \guess over GRUs-based baseline methods indicates that 
on sparse recommendation datasets, 
our simple attentive method could be easier to learn well than the 
complicated GRUs-based methods
and thus, enable better performance.


\subsection{Comparison with Graph-based Methods}
\label{sec:exp:gnn}

Graph-based methods~\cite{wu2019session,chen2020handling,xia2020self,qiu2019rethinking} are extensively developed for session-based recommendation, 
and have been demonstrated the state-of-the-art performance.
However, as shown in Table~\ref{tbl:performance}, 
\mjoint as a sequence-based method, significantly outperforms the state-of-the-art graph-based methods 
(i.e., \SRGNN, \LESSR, \DHCN) 
on the benchmark datasets. 
For example, \mjoint achieves significant improvement compared to the state-of-the-art graph-based methods at 
both recall@10 and recall@20 on all the datasets. 
Graph-based methods convert sessions to directed graphs or hyper graphs, 
and learn the complex temporal patterns~\cite{wu2019session} leveraging the graph structure (i.e., topology).
%
However, the graphs are constructed based on some assumptions by design
such as an item should link to all the subsequent items~\cite{chen2020handling}. 
Such assumptions may introduce noises or unnecessary/unrealistic relations in the graphs. Meanwhile, the sparse nature 
of recommendation datasets, and thus of the graphs, may not support the complicated learning of GNN-based models very well.
%
%
The superior performance of \mjoint over graph-based methods signifies that 
the sequence representation could be more effective than graphs for the recommendation.


%

\subsection{Analysis on Position Embeddings}
\label{sec:exp:abla}


\begin{table*}
  \caption{Performance Improvement of Position Embeddings}
  \label{tbl:abla}
  \centering
  \begin{threeparttable}
      \begin{scriptsize}
      \begin{tabular}{
	@{\hspace{0pt}}l@{\hspace{2pt}}
	@{\hspace{2pt}}l@{\hspace{2pt}}          
	@{\hspace{2pt}}r@{\hspace{2pt}}
        @{\hspace{2pt}}r@{\hspace{2pt}}
        @{\hspace{2pt}}r@{\hspace{2pt}}
        @{\hspace{1pt}}r@{\hspace{1pt}}
        @{\hspace{2pt}}r@{\hspace{2pt}}
        @{\hspace{2pt}}r@{\hspace{2pt}}
        @{\hspace{2pt}}r@{\hspace{2pt}}
        @{\hspace{1pt}}r@{\hspace{1pt}}
        @{\hspace{2pt}}r@{\hspace{2pt}}
        @{\hspace{2pt}}r@{\hspace{2pt}}
        @{\hspace{2pt}}r@{\hspace{2pt}}
        @{\hspace{0pt}}l@{\hspace{2pt}}
	@{\hspace{2pt}}l@{\hspace{2pt}}          
	@{\hspace{2pt}}r@{\hspace{2pt}}
        @{\hspace{2pt}}r@{\hspace{2pt}}
        @{\hspace{2pt}}r@{\hspace{2pt}}
        @{\hspace{1pt}}r@{\hspace{1pt}}
        @{\hspace{2pt}}r@{\hspace{2pt}}
        @{\hspace{2pt}}r@{\hspace{2pt}}
        @{\hspace{2pt}}r@{\hspace{2pt}}
        @{\hspace{1pt}}r@{\hspace{1pt}}
        @{\hspace{2pt}}r@{\hspace{2pt}}
        @{\hspace{2pt}}r@{\hspace{2pt}}
        @{\hspace{2pt}}r@{\hspace{2pt}}
	}
        \toprule
        \multirow{2}{*}{method} & \multirow{2}{*}{} & \multicolumn{3}{c}{recall@$k$} 
        && \multicolumn{3}{c}{MRR@$k$} && \multicolumn{3}{c}{NDCG@$k$}
        &&& \multicolumn{3}{c}{recall@$k$} 
        && \multicolumn{3}{c}{MRR@$k$} && \multicolumn{3}{c}{NDCG@$k$}\\
        \cline{3-5} \cline{7-9} \cline{11-13}
        \cline{16-18} \cline{20-22} \cline{24-26}
        & & $k$=5 & $k$=10 & $k$=20 && $k$=5 & $k$=10 & $k$=20 && $k$=5 & $k$=10 & $k$=20
        &&&$k$=5 & $k$=10 & $k$=20 && $k$=5 & $k$=10 & $k$=20 && $k$=5 & $k$=10 & $k$=20\\
        \midrule
        \mbox{\guess\!\textbackslash\pe}
        & \multirow{4}{*}{\rotatebox[origin=c]{90}{\DG}}
        & 0.2766 & 0.3918 & 0.5236 && 0.1580 & 0.1733 & 0.1824 && 0.1873 & 0.2245 & 0.2578
        && \multirow{4}{*}{\rotatebox[origin=c]{90}{\YC}}
        & 0.4138 & 0.5521 & 0.6754 && 0.2418 & 0.2603 & 0.2690 && 0.2845 & 0.3293 & 0.3606\\
        \guess && \textbf{0.2936} & \textbf{0.4148} & \textbf{0.5500} && \textbf{0.1657} & \textbf{0.1817} & \textbf{0.1911} && \textbf{0.1973} & \textbf{0.2364} & \textbf{0.2705}
        &&& \textbf{0.4721} & \textbf{0.6118} & \textbf{0.7203} && \textbf{0.2767} & \textbf{0.2956} & \textbf{0.3033} && \textbf{0.3254} & \textbf{0.3707} & \textbf{0.3983}\\
        \cline{3-13} \cline{16-26}
        \mbox{\mjoint\!\textbackslash\pe}
        && 0.2765 & 0.3934 & 0.5276 && 0.1570 & 0.1725 & 0.1818 && 0.1865 & 0.2242 & 0.2581
        &&& 0.4270 & 0.5674 & 0.6896 && 0.2470 & 0.2658 & 0.2744 && 0.2917 & 0.3371 & 0.3682\\
        \mjoint && \textbf{0.2919} & \textbf{0.4125} & \textbf{0.5476} && \textbf{0.1652} & \textbf{0.1812} & \textbf{0.1905} && \textbf{0.1966} & \textbf{0.2354} & \textbf{0.2695} 
        &&& \textbf{0.4834} & \textbf{0.6192} & \textbf{0.7277} && \textbf{0.2825} & \textbf{0.3008} & \textbf{0.3085}  && \textbf{0.3325} & \textbf{0.3766} & \textbf{0.4043}\\
        \midrule
        \mbox{\guess\!\textbackslash\pe}
        & \multirow{4}{*}{\rotatebox[origin=c]{90}{\GA}}
        & 0.3610 & 0.4392 & 0.5189 && 0.2275 & 0.2379 & 0.2435 && 0.2609 & 0.2862 & 0.3064
        && \multirow{4}{*}{\rotatebox[origin=c]{90}{\LF}}
        & \textbf{0.1165} & \textbf{0.1686} & \textbf{0.2360} && 0.0656 & \textbf{0.0724} & \textbf{0.0771} && \textbf{0.0782} & \textbf{0.0949} & \textbf{0.1119}\\
        \guess && \textbf{0.3700} & \textbf{0.4529} & \textbf{0.5352} && \textbf{0.2273} & \textbf{0.2384} & \textbf{0.2441} && \textbf{0.2629} & \textbf{0.2898} & \textbf{0.3106}
        &&& 0.1142 & 0.1646 & 0.2301 && \textbf{0.0658} & \textbf{0.0724} & 0.0769 && 0.0778 & 0.0940 & 0.1105 \\
        \cline{3-13} \cline{16-26}
        \mbox{\mjoint\!\textbackslash\pe}
        && 0.3677 & 0.4472 & 0.5277 && 0.2319 & 0.2425 & 0.2481 && 0.2658 & 0.2916 & 0.3119
        &&& 0.1232 & 0.1768 & 0.2449 && 0.0687 & 0.0757 & 0.0804 && 0.0822 & 0.0994 &0.1166\\
        \mjoint && \textbf{0.3803} & \textbf{0.4644} & \textbf{0.5479} && \textbf{0.2473} & \textbf{0.2586} & \textbf{0.2644} && \textbf{0.2805} & \textbf{0.3077} & \textbf{0.3289}
        &&& \textbf{0.1247} & \textbf{0.1771} & \textbf{0.2454} && \textbf{0.0724} & \textbf{0.0793} & \textbf{0.0840} && \textbf{0.0854} & \textbf{0.1022} & \textbf{0.1195}\\
        \midrule
        \mbox{\guess\!\textbackslash\pe} & \multirow{4}{*}{\rotatebox[origin=c]{90}{\NP}}
        & 0.1034 & 0.1589 & 0.2224 && 0.0591 & 0.0664 & 0.0708 && 0.0700 & 0.0878 & 0.1039
        && \multirow{4}{*}{\rotatebox[origin=c]{90}{\TM}}
        & \textbf{0.2320} & \textbf{0.2872} & \textbf{0.3432} && \textbf{0.1561} & \textbf{0.1635} & \textbf{0.1675} && \textbf{0.1750} & \textbf{0.1929} & \textbf{0.2071}\\
        \guess && \textbf{0.1180} & \textbf{0.1744} & \textbf{0.2371} && \textbf{0.0665} & \textbf{0.0740} & \textbf{0.0783} && \textbf{0.0791} & \textbf{0.0973} & \textbf{0.1132}
        &&& 0.2279 & 0.2840 & 0.3406 && 0.1534 & 0.1608 & 0.1648 && 0.1719 & 0.1900 & 0.2043\\
        \cline{3-13} \cline{16-26}
        \mbox{\mjoint\!\textbackslash\pe}
        && 0.1058 & 0.1598 & 0.2205 && 0.0603 & 0.0674 & 0.0717 && 0.0715 & 0.0889 & 0.1042
        &&& \textbf{0.2277} & \textbf{0.2861} & \textbf{0.3468} && \textbf{0.1483} & \textbf{0.1561} & \textbf{0.1603} && \textbf{0.1680} & \textbf{0.1869} & \textbf{0.2023}\\
        \mjoint && \textbf{0.1158} & \textbf{0.1688} & \textbf{0.2315} && \textbf{0.0662} & \textbf{0.0732} & \textbf{0.0775} && \textbf{0.0784} & \textbf{0.0955} & \textbf{0.1113}
        &&& 0.2192 & 0.2826 & 0.3438 && 0.1400 & 0.1484 & 0.1527 && 0.1597 & 0.1801 & 0.1956\\
        \bottomrule
      \end{tabular}
      \end{scriptsize}
      \begin{tablenotes}
        \setlength\labelsep{0pt}
	\begin{footnotesize}
	\item
        In this table, \mbox{\!\textbackslash\pe} represents \method variants (i.e., \guess and \mjoint) without position embeddings. 
        The best performance between \method variants with and without position embeddings (i.e. \mbox{\!\textbackslash\pe}) is in \textbf{bold}.
        \par
	\end{footnotesize}
      \end{tablenotes}
  \end{threeparttable}
\end{table*}


We conduct an analysis to verify the importance of position embeddings in \method. 
Specifically, in \method, 
we remove the position embeddings (i.e., \pos) in Equation~\ref{eqn:dot} and Equation~\ref{eqn:multi}, 
and calculate the attention weights using item embeddings (i.e., \emb) only.
We denote \method without position embeddings as \mbox{\method\!\textbackslash\pe}, 
and report the performance of \method and \mbox{\method\!\textbackslash\pe} in Table~\ref{tbl:abla}.
Due to the space limit, we do not present the performance of \msep
but we observed a similar trend in \msep.
%
%

As presented in Table~\ref{tbl:abla}, 
without position embeddings, the performance of \mbox{\guess\!\textbackslash\pe} and \mbox{\mjoint\!\textbackslash\pe} degrades significantly 
on four out of six datasets (i.e., \DG, \YC, \GA and \NP). 
For example, on \DG and \YC, \mbox{\mjoint\!\textbackslash\pe} underperforms \mjoint
at 3.8\% and 5.5\%, respectively. 
Recall that in \method, we learn position embeddings to incorporate the position information into the model, and better model the temporal patterns.
These results demonstrate the importance of the position information for session-based recommendation.
These results also reveal that the temporal patterns on the four datasets (e.g., \DG and \NP) are strong, 
and explain the similar performance of \guess and \mjoint on \DG and \NP as discussed in Section~\ref{sec:exp:variant}.
We also notice that on \LF and \TM, \mbox{\guess\!\textbackslash\pe} and \mbox{\mjoint\!\textbackslash\pe} still
achieve performance similar to that with position embeddings.
For example, on \LF, \mbox{\mjoint\!\textbackslash\pe} achieves 0.2449 at recall@20, 
and \mjoint achieves 0.2454 (difference: 0.2\%). 
Similarly, on \TM, at recall@20, the performance of \mbox{\mjoint\!\textbackslash\pe} and \mjoint is 0.3468 and 0.3438, respectively (difference: 0.9\%).
%
As will be shown in Section~\ref{sec:exp:att}, on some datasets (e.g., \LF), the position information may not be crucial for the recommendation.
Therefore, on these datasets, without position embeddings, the model could still achieve similar performance.

\subsection{\mbox{Prospective Preference Estimate Analysis}}
\label{sec:exp:global}

\begin{table}
  \caption{Performance Comparison on Estimate Strategies}
  \label{tbl:guess}
  \centering
  \begin{threeparttable}
      \begin{scriptsize}
      \begin{tabular}{
	    @{\hspace{0pt}}l@{\hspace{2pt}}
	    @{\hspace{2pt}}l@{\hspace{2pt}}          
	    @{\hspace{2pt}}r@{\hspace{2pt}}
        @{\hspace{2pt}}r@{\hspace{2pt}}
        @{\hspace{2pt}}r@{\hspace{2pt}}
        @{\hspace{2pt}}r@{\hspace{2pt}}
        @{\hspace{2pt}}r@{\hspace{2pt}}
        @{\hspace{2pt}}r@{\hspace{2pt}}
        @{\hspace{2pt}}r@{\hspace{2pt}}
        @{\hspace{2pt}}r@{\hspace{2pt}}
        @{\hspace{2pt}}r@{\hspace{2pt}}
        @{\hspace{2pt}}r@{\hspace{2pt}}
        @{\hspace{2pt}}r@{\hspace{0pt}}
	}
        \toprule
        \multirow{2}{*}{} & \multirow{2}{*}{method} & \multicolumn{3}{c}{recall@$k$} 
        && \multicolumn{3}{c}{MRR@$k$} && \multicolumn{3}{c}{NDCG@$k$}\\
        \cline{3-5} \cline{7-9} \cline{11-13}
        & & $k$=5 & $k$=10 & $k$=20 && $k$=5 & $k$=10 & $k$=20 && $k$=5 & $k$=10 & $k$=20\\
        \midrule
        \multirow{2}{*}{\rotatebox[origin=c]{90}{\DG}}
        & \ljoint & 0.2881 & 0.4098 & 0.5459 && 0.1615 & 0.1776 & 0.1870 && 0.1928 & 0.2320 & 0.2664\\
       & \mjoint & \textbf{0.2922} & \textbf{0.4120} & \textbf{0.5474} && \textbf{0.1646} & \textbf{0.1805} & \textbf{0.1899} 
       && \textbf{0.1961} & \textbf{0.2348} & \textbf{0.2690} \\
        \midrule
        \multirow{2}{*}{\rotatebox[origin=c]{90}{\YC}}
        & \ljoint & \textbf{0.4851} & \textbf{0.6202} & 0.7265 && 0.2824 & 0.3006 & 0.3081 && \textbf{0.3329} & \textbf{0.3768} & 0.4038\\
        & \mjoint & 0.4834 & 0.6192 & \textbf{0.7277} && \textbf{0.2825} & \textbf{0.3008} & \textbf{0.3085} && 0.3325 & 0.3766 & \textbf{0.4043}\\
        \midrule
        \multirow{2}{*}{\rotatebox[origin=c]{90}{\GA}}
        & \ljoint & 0.3777 & 0.4638 & 0.5475 && 0.2339 & 0.2454 & 0.2512 && 0.2698 & 0.2977 & 0.3189\\
        & \mjoint & \textbf{0.3803} & \textbf{0.4644} & \textbf{0.5479} && \textbf{0.2473} & \textbf{0.2586} & \textbf{0.2644} && \textbf{0.2805} & \textbf{0.3077} & \textbf{0.3289}\\
        \midrule
        \multirow{2}{*}{\rotatebox[origin=c]{90}{\LF}}
        & \ljoint & 0.1176 & 0.1676 & 0.2332 && 0.0681 & 0.0747 & 0.0792 && 0.0804 & 0.0965 & 0.1130\\
        & \mjoint & \textbf{0.1247} & \textbf{0.1771} & \textbf{0.2454} && \textbf{0.0724} & \textbf{0.0793} & \textbf{0.0840} && \textbf{0.0854} & \textbf{0.1022} & \textbf{0.1195}\\
        \midrule
        \multirow{2}{*}{\rotatebox[origin=c]{90}{\NP}}
        & \ljoint & \textbf{0.1261} & \textbf{0.1735} & 0.2304 && \textbf{0.0736} & \textbf{0.0799} & \textbf{0.0838} && \textbf{0.0866} & \textbf{0.1019} & \textbf{0.1163}\\
        & \mjoint & 0.1158 & 0.1688 & \textbf{0.2315} && 0.0662 & 0.0732 & 0.0775 && 0.0784 & 0.0955 & 0.1113\\
        \midrule
        \multirow{2}{*}{\rotatebox[origin=c]{90}{\TM}}
        & \ljoint & 0.2071 & 0.2637 & 0.3176 && 0.1324 & 0.1401 & 0.1439 && 0.1510 & 0.1694 & 0.1830\\
        & \mjoint  & \textbf{0.2192} & \textbf{0.2826} & \textbf{0.3438} && \textbf{0.1400} & \textbf{0.1484} & \textbf{0.1527} && \textbf{0.1597} & \textbf{0.1801} & \textbf{0.1956}\\
        \bottomrule
      \end{tabular}
      \end{scriptsize}
      \begin{tablenotes}
      \setlength\labelsep{0pt}
	\begin{footnotesize}
	\item
	In this table, \ljoint estimates users' prospective preferences using the last item in each session, 
	and \mjoint estimates users' prospective preferences using \ppp (Section~\ref{sec:method:predict}).
	The best performance in each dataset is in \textbf{bold}. 
        \par
	\end{footnotesize}
      \end{tablenotes}
  \end{threeparttable}
\end{table}

%
%
%

In \method, we leverage the position-sensitive preference prediction (i.e., $\pref$) from \ppp (Section~\ref{sec:method:predict}) as the estimate
of users' prospective preferences (Section~\ref{sec:method:items}).
%
We notice that in the literature~\cite{liu2018stamp,wu2019session,kang2018self}, 
another strategy is to estimate the users' prospective preferences from the last item in each session.
This strategy is based on the recency assumption~\cite{liu2018stamp,wu2019session} that 
the most recently interacted item could be a highly strong indicator of the next item of users' interest. 
%
We conduct an analysis to empirically compare the two strategies in \mjoint. 
%
Specifically, in Equation~\ref{eqn:multi}, instead of $\pref$,
we use the embeddings of the last item and its position in each session (i.e., $\textbf{v}_{a_\nn}\!+\mathbf{p}_\nn$) to calculate the attention weights, 
and denote the resulted variant as \ljoint. 
%
Table~\ref{tbl:guess} presents the performance of \mjoint and \ljoint.
Similarly to that in \mjoint, we tune hyper parameters for \ljoint using grid search, and report the results from the identified best performing hyper parameters.
%

As presented in Table~\ref{tbl:guess}, 
overall \mjoint achieves considerable improvement compared to \ljoint on four out of six datasets (i.e., \DG, \GA, \LF and \TM). 
At recall@5, recall@10 and recall@20, on average, \mjoint achieves significant improvement of 3.5\%, 3.4\% and 3.5\%, respectively, over \ljoint.
We find a similar trend at MRR@$k$ and NDCG@$k$.
For example, in terms of MRR@5 and NDCG@5, over the four datasets, 
\mjoint still significantly outperforms \ljoint with an average improvement of 4.9\% and 4.4\%, respectively.
The primary difference between \mjoint and \ljoint is that \mjoint uses a learning-based method to estimate users' prospective preferences in a data-driven manner, 
while \ljoint generates the estimate based on the recency assumption.
The above results indicate that on most of the datasets, 
data driven-based estimate is more effective than recency-based estimate.
We also notice that
on \YC and \NP, the performance of \mjoint is competitive with that of \ljoint.
For example, on \YC, the performance difference between \mjoint and \ljoint is 0.2\% and 0.1\% at recall@20 and MRR@20, respectively.
On \NP, the difference at recall@20 is also as small as 0.5\%.
As shown in the literature~\cite{liu2018stamp}, some datasets such as \YC have strong recency patterns. 
%
The similar results between \mjoint and \ljoint on \YC and \NP, 
indicate that on datasets with strong recency patterns, 
our learning-based strategy may implicitly capture the patterns, and still produce competitive results.
%

\subsection{Attention Weight Analysis}
\label{sec:exp:att}

\begin{figure}[!h]
       \centering
       \footnotesize
        \begin{minipage}{\linewidth}
               \begin{subfigure}{0.45\linewidth}
                    \centering
                    \hspace*{20pt}
                    \input{attention_YC_ppp2}
                    \vspace*{-10pt}
                    \caption{\YC (\ppp)}
                    \label{fig:yc_ppp}
                \end{subfigure}
%
%
                \begin{subfigure}{0.45\linewidth}
                    \centering
                    \hspace*{20pt}                    
                    \input{attention_YC_ppp_pos2}
                    \vspace*{-10pt}                    
                    \caption{\YC (\mbox{\ppp\!\textbackslash\pe})}
                    \label{fig:yc_ppp_pos}
                \end{subfigure}
\vspace{-10pt}
        \end{minipage}
%
        \begin{minipage}{\linewidth}
               \begin{subfigure}{0.45\linewidth}
                    \centering
                    \hspace*{20pt}
                    \input{attention_LF_ppp2}
                    \vspace*{-10pt}
                    \caption{\LF (\ppp)}
                    \label{fig:lf_ppp}
                \end{subfigure}
                \begin{subfigure}{0.45\linewidth}
                    \centering
                    \hspace*{20pt}                    
                    \input{attention_LF_ppp_pos2}
                    \vspace*{-10pt}
                    \caption{\LF (\mbox{\ppp\!\textbackslash\pe})}
                    \label{fig:lf_ppp_pos}
                \end{subfigure}
%

          \end{minipage}
%
\caption{Attention Weights from \mjoint}
\label{fig:att}
\end{figure}

We conduct an analysis to evaluate the attention weights learned in \mjoint.
Particularly, we represent the attention weights learned in \ppp (Section~\ref{sec:method:predict}) over corresponding session 
positions in Figure~\ref{fig:att}.
Due to the space limit we only present the results on \YC and \LF 
but we have similar results on the other datasets.
%
%
In Figure~\ref{fig:att}, the $y$-axis corresponds to the session lengths; the $x$-axis corresponds to the 
last $n$ positions of a session of length $n$ ($n=1, \cdots, 8$).
%
Due to the space limit we only present the weights in augmented testing sessions with at most 8 items, 
which represent 82.2\% and 57.4\% of all the testing sessions in \YC and \LF, respectively.
However, we observed a similar trend in the longer sessions as that in Figure~\ref{fig:att}.
%

Figure~\ref{fig:yc_ppp} and \ref{fig:yc_ppp_pos} present the attention weight distribution from 
\ppp with, and without position embeddings (i.e., \mbox{\!\textbackslash\pe}), respectively, on \YC.
Similarly, Figure~\ref{fig:lf_ppp} and \ref{fig:lf_ppp_pos} present the distributions on \LF.
Comparing Figure ~\ref{fig:yc_ppp} and Figure~\ref{fig:yc_ppp_pos}, 
we notice that on \YC, the attention weights on the last item of sessions (i.e., diagonal in the Figure)
are significantly higher 
than those on the earlier ones. 
%
%
This weight distribution is consistent with the recency pattern on \YC 
as shown in the literature~\cite{liu2018stamp}, 
and demonstrates that with the position embeddings, 
\ppp could accurately capture the recency patterns in \YC.
%
Figure~\ref{fig:yc_ppp_pos} shows that without position embeddings, 
the attention weights from \ppp do not show considerable difference over positions, 
revealing that without position embeddings, the attention mechanism cannot differentiate positions..
%
The comparison between Figure~{\ref{fig:yc_ppp}} and Figure~\ref{fig:yc_ppp_pos} further demonstrates 
the importance of position embeddings in \method.

Comparing Figure~\ref{fig:yc_ppp} and Figure~\ref{fig:lf_ppp}, 
we find that on \YC, the weights follow the recency pattern, 
while on \LF, 
the second last item has higher weights than the last one.
This result implies that the recency assumption may not always hold, 
and our learning-based method (i.e., \ppp) could be more effective 
than the existing recency-based method in estimating prospective preferences (Section~\ref{sec:exp:global}).
Comparing Figure~\ref{fig:lf_ppp} and Figure~\ref{fig:lf_ppp_pos}, 
we notice that without position embeddings, on sessions with more than 4 items, the weight distribution in Figure~\ref{fig:lf_ppp_pos} 
is still similar with that in Figure~\ref{fig:lf_ppp}.
%
This result reveals that on this dataset, the position information may not be critical, 
and also supports the similar performance of \method and \mbox{\method\!\textbackslash\pe} on \LF as discussed in Section~\ref{sec:exp:abla}.
\subsection{Analysis on Cosine Similarities}
\label{sec:exp:sim}

\begin{table}[!h]
  \caption{Preference Prediction Similarity Comparison}
  \label{tbl:sim}
  \centering
  \begin{threeparttable}
      \begin{footnotesize}
      \begin{tabular}{
	@{\hspace{6pt}}l@{\hspace{6pt}}
        @{\hspace{6pt}}r@{\hspace{6pt}}
        @{\hspace{6pt}}r@{\hspace{6pt}}
        @{\hspace{6pt}}r@{\hspace{6pt}}
        @{\hspace{6pt}}r@{\hspace{6pt}}
        @{\hspace{6pt}}r@{\hspace{6pt}}
        %
        %
        @{\hspace{6pt}}r@{\hspace{6pt}}
        @{\hspace{6pt}}r@{\hspace{6pt}}
        @{\hspace{6pt}}r@{\hspace{6pt}}
        @{\hspace{6pt}}r@{\hspace{6pt}}
        @{\hspace{6pt}}r@{\hspace{6pt}}
        @{\hspace{6pt}}r@{\hspace{0pt}}
	}
        \toprule
        %
        %
        %
        \multirow{2}{*} {dataset} & \multicolumn{2}{c}{\guess} && \multicolumn{2}{c}{\msep} && \multicolumn{2}{c}{\mjoint} \\
        \cline{2-3} \cline{5-6} \cline{8-9}
        & avg. & next && avg. & next && avg. & next\\
        \midrule
        \DG & -0.050 & 0.662 && -0.007 & 0.646 && -0.010 & 0.661\\
        \YC & -0.137 & 0.724 && -0.084 & 0.660 && -0.082 & 0.666\\
        \GA & -0.030 & 0.676 && -0.007 & 0.663 && -0.001 & 0.610\\
        \LF & -0.150 & 0.322 && -0.006 & 0.325 && 0.012 & 0.306\\
        \NP & -0.057 & 0.423 && 0.014 & 0.454 && 0.007 & 0.422\\
        \TM & -0.042 & 0.356 && 0.002 & 0.383 && 0.015 & 0.370\\
        \bottomrule
      \end{tabular}
      \end{footnotesize}
      \begin{tablenotes}
        \setlength\labelsep{0pt}
	\begin{footnotesize}
	\item
	In this table, the columns avg. shows the average cosine similarities between preference predictions and all the candidate items, 
	The columns next shows the similarity between the predictions and the ground-truth next item.
        \par
	\end{footnotesize}
      \end{tablenotes}
  \end{threeparttable}
\end{table}


We conduct an analysis to verify if \method truly learns predictive preferences from the data.
Specifically, we calculate the cosine similarities between the preference predictions (i.e., $\pref$ and $\nei$) and the item embeddings on the six datasets, 
and present the results in Table~\ref{tbl:sim}.
For \guess/\msep, we calculate similarities between $\pref$/$\nei$ and item embeddings.
For \mjoint, since we calculate recommendation scores using $\pref$+$\nei$ (Equation~\ref{eqn:score:joint}), 
we use $\pref$+$\nei$ to calculate the similarities.
%
%
%
Table~\ref{tbl:sim} shows that in \guess, \msep and \mjoint, 
compared to the average similarities among all the items, 
the similarities between the predictions and the ground-truth next item is significantly higher.
This result reveals that \method could learn to capture users' true preferences from the data.

\subsection{Run-time Performance}
\label{sec:exp:runtime}

\begin{table}[!h]
  \caption{Testing Runtime Performance (ms)}
  \label{tbl:runtime}
  \centering
  \begin{threeparttable}
      \begin{footnotesize}
      \begin{tabular}{
	@{\hspace{0pt}}l@{\hspace{8pt}}
        @{\hspace{8pt}}r@{\hspace{8pt}}
        @{\hspace{8pt}}r@{\hspace{8pt}}
        @{\hspace{8pt}}r@{\hspace{8pt}}
        @{\hspace{8pt}}r@{\hspace{8pt}}
        @{\hspace{8pt}}r@{\hspace{8pt}}
        @{\hspace{8pt}}r@{\hspace{0pt}}
	}
        \toprule
        method & \DG & \YC & \GA & \LF & \NP & \TM\\
        \midrule
        \DHCN & 4.6e0 & 2.8e0 & 3.0e0 &1.3e2 & 7.2e0 & 4.4e0\\
        \mjoint & \textbf{5.7e-1} & \textbf{2.7e-1} & \textbf{4.2e-1} & \textbf{5.2e-1} & \textbf{8.9e-1} & \textbf{6.2e-1}\\
        \cmidrule(lr){1-7}
        speedup & 8.1 & 10.4 & 7.1 & 245.4 & 8.1 & 7.1\\
        \bottomrule
      \end{tabular}
      \end{footnotesize}
      \begin{tablenotes}
       \setlength\labelsep{0pt}
	\begin{footnotesize}
	\item
	The best run-time performance at each dataset is in \textbf{bold}.
        \par
	\end{footnotesize}
      \end{tablenotes}
  \end{threeparttable}
\end{table}


We compare the run-time performance of \mjoint and that of the
best performing baseline method \DHCN during testing, 
and report the results in Table~\ref{tbl:runtime}.
We focus on testing instead of training due to the fact that 
compared to training, 
the run-time performance in testing could better 
imply the models' latency in real-time recommendation, 
which could significantly affect the user experience and thus revenue.
%
%
As presented in Table~\ref{tbl:runtime}, the run-time performance of {\mjoint} is substantially better than that of \DHCN 
on all the datasets.
%
Specifically, on average, \mjoint is 47.7 times faster than \DHCN.
The superior run-time performance of {\mjoint} over \DHCN 
demonstrates that while generating high-quality recommendations, 
{\method} could enable lower latency in real time, 
and thus could significantly improve the user experience.


\subsection{Parameter Study}
\label{sec:exp:para}

\begin{figure}[!h]
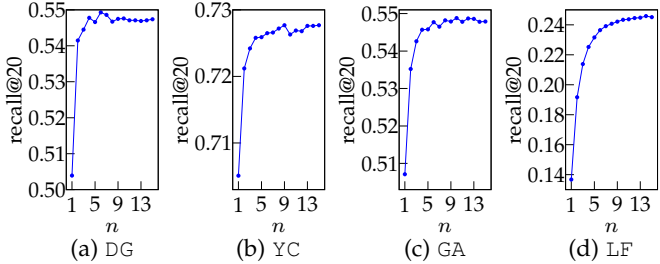

      \centering
    \footnotesize
    
        
%
                \begin{subfigure}{0.24\linewidth}
                    \centering
                        \input{DG2.tex}
                        \vspace{-10pt}
                        \hspace{-10pt}
                        \caption{\DG}\label{fig:para:dg}
                \end{subfigure}
                 \begin{subfigure}{0.24\linewidth}
                    \centering
                        \input{YC2.tex}
                        \vspace{-10pt}
                        \hspace{-10pt}
                         \caption{\YC}\label{fig:para:yc}
                \end{subfigure}
                 \begin{subfigure}{0.24\linewidth}
                    \centering
                        \input{GA2.tex}
                        \vspace{-10pt}
                        \hspace{-10pt}    
                        \caption{\GA}\label{fig:para:ga}
                \end{subfigure}
                 \begin{subfigure}{0.24\linewidth}
                    \centering
                        \input{LF2.tex}
                          \vspace{-10pt}
                          \hspace{-10pt}
                        \caption{\LF}\label{fig:para:lf}
                \end{subfigure}
                %
\caption{Parameter Study}
\label{fig:para}
\end{figure}

We conduct a parameter study to assess how the length of the transformed sequence (i.e., $\nn$) 
affect the recommendation performance on the widely used \DG, \YC, \GA and \LF datasets.
Particularly, on each dataset, we change $\nn$ and fix the other hyper parameters as the best performing ones during hyper parameter tuning,
and report the performance at recall@20 on augmented testing sessions in Figure~\ref{fig:para}.
As shown in Figure~\ref{fig:para}, on \DG, \YC and \GA, the performance increases significantly as $\nn$ increases when $\nn < 5$, 
while when $\nn \geq 5$, incorporating earlier items in the session will not considerably improve the performance.
Similarly on \LF, the performance becomes stable when $\nn \geq 10$. 
These results reveal that on session-based recommendation datasets, 
only the most recent few items are effective in learning users' preferences.
As a result, we will not loss crucial information in \method by using only the last $\nn$ items (Section~\ref{sec:method:embed}) for the recommendation.


\section{Discussion}
\label{sec:dis}


\begin{table}
  \caption{Effectiveness of Users' Prospective Preferences}
  \label{tbl:oracle}
  \centering
  \begin{threeparttable}
      \begin{scriptsize}
      \begin{tabular}{
	@{\hspace{0pt}}l@{\hspace{2pt}}
	@{\hspace{2pt}}l@{\hspace{2pt}}
	@{\hspace{2pt}}r@{\hspace{2pt}}
        @{\hspace{2pt}}r@{\hspace{2pt}}
        @{\hspace{2pt}}r@{\hspace{2pt}}
        @{\hspace{2pt}}r@{\hspace{2pt}}
        @{\hspace{2pt}}r@{\hspace{2pt}}
        @{\hspace{2pt}}r@{\hspace{2pt}}
        @{\hspace{2pt}}r@{\hspace{2pt}}
        @{\hspace{2pt}}r@{\hspace{2pt}}
        @{\hspace{2pt}}r@{\hspace{2pt}}
        @{\hspace{2pt}}r@{\hspace{2pt}}
        @{\hspace{2pt}}r@{\hspace{0pt}}
	}
        \toprule
        \multirow{2}{*}{} & \multirow{2}{*}{method} & \multicolumn{3}{c}{recall@$k$} 
        && \multicolumn{3}{c}{MRR@$k$} && \multicolumn{3}{c}{NDCG@$k$}\\
        \cline{3-5} \cline{7-9} \cline{11-13}
        & & $k$=5 & $k$=10 & $k$=20 && $k$=5 & $k$=10 & $k$=20 && $k$=5 & $k$=10 & $k$=20\\
        \midrule
        \multirow{2}{*}{\rotatebox[origin=c]{90}{\DG}}
        & \MEAN  & 0.2778 & 0.3917 & 0.5246 && 0.1585 & 0.1736 & 0.1828 && 0.1880 & 0.2247 & 0.2583\\
        & \ORACLE & \textbf{0.4113} & \textbf{0.5260} & \textbf{0.6436} 
        && \textbf{0.2755} & \textbf{0.2908} & \textbf{0.2990} 
        && \textbf{0.3092} & \textbf{0.3463} & \textbf{0.3760}\\
        \midrule
        \multirow{2}{*}{\rotatebox[origin=c]{90}{\YC}}
        & \MEAN  & 0.4039 & 0.5392 & 0.6620 && 0.2376 & 0.2557 & 0.2643 && 0.2788 & 0.3227 & 0.3538\\
        & \ORACLE & \textbf{0.5771} & \textbf{0.6922} & \textbf{0.7760} 
        && \textbf{0.3791} & \textbf{0.3947} & \textbf{0.4006} 
        && \textbf{0.4285} & \textbf{0.4660} & \textbf{0.4873}\\
        \midrule
        \multirow{2}{*}{\rotatebox[origin=c]{90}{\GA}}
        & \MEAN  & 0.3589 & 0.4387 & 0.5202 && 0.2243 & 0.2350 & 0.2407 && 0.2580 & 0.2838 & 0.3044\\
        & \ORACLE & \textbf{0.4350} & \textbf{0.5097} & \textbf{0.5837} 
        && \textbf{0.3084} & \textbf{0.3185} & \textbf{0.3236} 
        && \textbf{0.3401} & \textbf{0.3643} & \textbf{0.3830}\\
        \midrule
        \multirow{2}{*}{\rotatebox[origin=c]{90}{\LF}}
        & \MEAN  & 0.1200 & 0.1706 & 0.2366 && 0.0685 & 0.0751 & 0.0797 && 0.0812 & 0.0975 & 0.1142\\
        & \ORACLE & \textbf{0.2544} & \textbf{0.3152} & \textbf{0.3892} 
        && \textbf{0.1915} & \textbf{0.1996} & \textbf{0.2047} 
        && \textbf{0.2071} & \textbf{0.2267} & \textbf{0.2453}\\
        \midrule
        \multirow{2}{*}{\rotatebox[origin=c]{90}{\NP}}
        & \MEAN & 0.1019 & 0.1592 & 0.2250 && 0.0579 & 0.0654 & 0.0700 && 0.0687 & 0.0871 & 0.1037\\
        & \ORACLE & \textbf{0.1822} & \textbf{0.2429} & \textbf{0.3088} 
        && \textbf{0.1078} & \textbf{0.1160} & \textbf{0.1205} 
        && \textbf{0.1263} & \textbf{0.1459} & \textbf{0.1625}\\
        \midrule
        \multirow{2}{*}{\rotatebox[origin=c]{90}{\TM}}
        & \MEAN & 0.2531 & 0.3181 & 0.3780 && 0.1676 & 0.1763 & 0.1805 && 0.1889 & 0.2099 & 0.2251\\
        & \ORACLE & \textbf{0.5282} & \textbf{0.5717} & \textbf{0.6086} 
        && \textbf{0.4489} & \textbf{0.4547} & \textbf{0.4573} 
        && \textbf{0.4687} & \textbf{0.4828} & \textbf{0.4921}\\
        \bottomrule
      \end{tabular}
      \end{scriptsize}
      \begin{tablenotes}
      \setlength\labelsep{0pt}
	\begin{footnotesize}
	\item
	In this table, \MEAN equally weighs items in the session, 
	and \ORACLE learns attention weights over items conditioned on the ground-truth next item.
        The best performance at each dataset is in \textbf{bold}.
        \par
	\end{footnotesize}
      \end{tablenotes}
  \end{threeparttable}
\end{table}


We conduct an experiment to verify that users' prospective preferences could signify the important items, 
and thus, benefit the recommendation.
Specifically, we develop a method, denoted as \ORACLE, 
which learns attention weights conditioned on the ground-truth next item (i.e,,  $s_{|S|+1}$, the true preference).
%
In \ORACLE, we generate recommendations in the same way as that in \guess 
except that in the dot-product attention (Equation~\ref{eqn:dot}), we remove the position embeddings (i.e., $\pos$) and replace $\Q$ with 
$\mathbf{v_{s_{|S|+1}}}$. (i.e., embedding of $s_{|S|+1}$).
%
We empirically compare \ORACLE with another method, denoted as \MEAN, which uses a mean pooling to equally weigh items in the session.
We report the results of \ORACLE and \MEAN on the six datasets in Table~\ref{tbl:oracle}.
%

As shown in Table~\ref{tbl:oracle}, \ORACLE significantly outperforms \MEAN in all the datasets.
For example, in terms of recall@5, MRR@5 and NDCG@5, compared to \MEAN, on average, \ORACLE achieves significant improvement of 68.6\%, 100.7\% and 89.5\%, respectively.
%
%
%
The superior performance of \ORACLE over \MEAN shows that the learned attention weights in \ORACLE are effective, 
and further reveals that users' prospective preferences could indicate important items.
These results motivate us to estimate the prospective preferences, 
and weigh items conditioned on the estimate as in \spp (Section~\ref{sec:method:items}).

%

We notice that on \TM, \MEAN outperforms \method and all the baseline methods (Table~\ref{tbl:performance}).
Previous work~\cite{peng2021ham} suggests that on some extremely sparse recommendation datasets, 
the attention-based methods may not be well-learned, and underperform simple mean pooling-based method (i.e., \MEAN).
\TM with the smallest training set and a large number of items is extremely sparse.
Therefore, \MEAN could outperform \method and all the baseline methods on this dataset.
However, on all the other datasets, \method significantly outperforms \MEAN, 
which reveals that on most of the datasets, 
\method could learn well, and thus, enable better performance.

\section{Conclusions}
\label{sec:con}

In this manuscript, we presented novel \method models 
that conduct session-based recommendations 
using two important factors: temporal patterns and estimates of users' prospective preferences.
Our experimental results in comparison with five state-of-the-art baseline methods
on the six benchmark datasets demonstrate that 
\method significantly outperforms the baseline methods with an improvement of up to 19.2\%.
The results also reveal that on most of the datasets, the two factors could reinforce each other, and enable superior performance.
Our analysis on position embeddings signifies the importance of explicitly modeling the position information for session-based recommendation.
Our analysis on 
prospective preference estimate strategies demonstrates that on most of the datasets, 
our learning-based strategy is more effective than the existing recency-based strategy.
Our analysis on the learned attention weights shows that with position embeddings, 
\method could effectively capture the temporal patterns (e.g., recency patterns).
Our results in run-time performance comparison show that \method is much 
more efficient than the best baseline method \DHCN (47.7 average speedup).
Our analysis on users' prospective preferences demonstrates that 
the prospective preferences could signify important items, and thus, benefit the recommendation.

\section*{Acknowledgement}
\label{app}

This project was made possible, in part, by support from the National Science Foundation 
under Grant Number IIS-1855501, EAR-1520870, SES-1949037, IIS-1827472 and IIS-2133650, 
and from National Library of Medicine under Grant Number 1R01LM012605-01A1 and R21LM013678-01. 
Any opinions, findings, and conclusions or recommendations expressed 
in this material are those of the authors and do not necessarily reflect the views of the funding agencies.

\appendices

\section{Reproducibility}
\label{app:rep}
%
%
\begin{table*}[h]
\footnotesize
  \caption{\mbox{Best Hyper Parameters for \method, \ljoint and Baseline Methods}}
  \centering
  \label{tbl:para}
  \begin{threeparttable}
      \begin{tabular}{
        @{\hspace{8pt}}l@{\hspace{8pt}}
        @{\hspace{2pt}}r@{\hspace{8pt}}
        @{\hspace{8pt}}r@{\hspace{8pt}}
        @{\hspace{2pt}}c@{\hspace{2pt}}
        @{\hspace{8pt}}r@{\hspace{8pt}}
        @{\hspace{8pt}}r@{\hspace{8pt}}
        @{\hspace{8pt}}r@{\hspace{2pt}}
	@{\hspace{2pt}}c@{\hspace{2pt}}
        @{\hspace{8pt}}r@{\hspace{8pt}}
        @{\hspace{8pt}}r@{\hspace{8pt}}
        @{\hspace{8pt}}r@{\hspace{2pt}}
	@{\hspace{2pt}}c@{\hspace{2pt}}
        @{\hspace{2pt}}r@{\hspace{8pt}}
        @{\hspace{8pt}}r@{\hspace{8pt}}
        @{\hspace{2pt}}c@{\hspace{2pt}}
        @{\hspace{2pt}}r@{\hspace{8pt}}
        @{\hspace{8pt}}r@{\hspace{8pt}}
        @{\hspace{8pt}}r@{\hspace{8pt}}
        @{\hspace{2pt}}c@{\hspace{2pt}}
        @{\hspace{2pt}}r@{\hspace{8pt}}
        @{\hspace{8pt}}r@{\hspace{5pt}}
        @{\hspace{8pt}}r@{\hspace{8pt}}
        @{\hspace{2pt}}c@{\hspace{2pt}}
        @{\hspace{2pt}}r@{\hspace{8pt}}
        @{\hspace{8pt}}r@{\hspace{8pt}}
        }
        \toprule
        \multirow{2}{*}{Dataset} & \multicolumn{2}{c}{\guess} && \multicolumn{3}{c}{\mjoint} && \multicolumn{3}{c}{\ljoint} && \multicolumn{2}{c}{\NARM} 
        && \multicolumn{3}{c}{\SRGNN} && \multicolumn{3}{c}{\LESSR} && \multicolumn{2}{c}{\DHCN}\\
        \cmidrule(lr){2-3} \cmidrule(lr){5-7} \cmidrule(lr){9-11} \cmidrule(lr){13-14} \cmidrule(lr){16-18} \cmidrule(lr){20-22} \cmidrule(lr){24-25}
        & $d$ & \nn
        && $d$ & \nn & \nh  
        && $d$ & \nn & \nh
        && $d$ & $lr$
        && $d$ & $lr$ & $\lambda$
        && $d$ & $lr$ & $l$
        && $d$ & $\beta$\\
        \midrule
        DG & 128 & 10 && 128 & 15 & 8 && 128 & 10 & 4 && 64 & 1e-3 && 32   & 1e-4 & 1e-6 && 32  & 2e-3 & 3 && 128 & 5e-3\\
        YC & 128 & 7   && 128 & 8   & 8 && 128 & 20 & 2 && 64  & 1e-3 && 128 & 1e-3  & 0.0  && 32  & 1e-3 & 2 && 128 & 1e-4\\        
        GA & 128 & 10 && 128 & 15 & 8 && 128 & 10 & 4 && 64  & 1e-3 && 128 & 1e-3 & 1e-6 && 32  & 1e-3 & 2 && 128 & 1e-3\\
        LF  & 128 & 10 && 128 & 20 & 2 && 128 & 20 & 1 &&128 & 1e-3 && 128 & 5e-4 & 0.0   && 128 & 1e-3 & 4 && 128 & 1e-5\\
        NP & 128 & 9   && 128 & 8   & 4 && 128 & 20 & 4 && 64  & 1e-3 && 128 & 5e-4 & 1e-6 && 32   & 2e-3 & 3 && 128 & 1e-3\\
        TM & 128 & 15 && 128 & 10 & 4 && 128 & 20 & 4 &&128 & 1e-4 && 96  & 1e-3 & 1e-6 && 128 & 5e-4 & 2 && 64 & 5e-5\\
        \bottomrule
      \end{tabular}
      \begin{tablenotes}
      \setlength\labelsep{0pt}
      \begin{footnotesize}
      \item
          In this table, in \method, $d$, $\nn$ and $\nh$ are
          the dimension of the hidden representation, length of the transformed session and the number of heads.
          In \NARM, \SRGNN and \LESSR, $d$ is the dimension of the hidden representation and $lr$ is the learning rate.
	  In \SRGNN, $\lambda$ is the weight decay factor.
	  In \LESSR, $l$ is the number of GNN layers, and in \DHCN, $\beta$ is the factor for the self-supervision. 
          \par
      \end{footnotesize}
      \end{tablenotes}
  \end{threeparttable}
\end{table*}

%
We implement \method in python 3.7.3 with PyTorch 1.4.0~\footnote{\url{https://pytorch.org}}.
We use Adam optimizer with learning rate 1e-3 on all the datasets for \method variants (i.e., \guess, \msep and \mjoint).
We initialize all the learnable parameters using the default initialization methods 
in PyTorch.
The source code and processed data is available on GitHub~\footnote{\url{https://github.com/ninglab/P2MAM}}. 
%
For all the methods, during the grid search, we initially search the hyper parameters in a search range. 
If the hyper parameter yields the best performance on the boundary of the search range,  
we will extend the search range, if applicable, until a value in the middle yields the best performance. 
Table~\ref{tbl:para} presents the hyper parameters used for all the methods. 

For \method and \ljoint, the initially search range 
for the embedding dimension $d$, the length of transformed sequence 
$\nn$ and the number of heads $\nh$ 
is $\{32, 64, 128\}$, $\{10, 15, 20\}$ and $\{1, 2, 4\}$, respectively.

For \NARM~\footnote{\mbox{\url{https://github.com/lijingsdu/sessionRec_NARM}}}, 
we initially search $d$ and the learning rate $lr$ from $\{32, 64, 128\}$ and $\{$1e-4, 1e-3$\}$, respectively.

For \SRGNN~\footnote{\mbox{\url{https://github.com/CRIPAC-DIG/SR-GNN}}}, 
we initially search $d$, $lr$ and the weight decay factor $\lambda$ from $\{32, 64, 128\}$, $\{$1e-4, 1e-3$\}$ and $\{$1e-5, 1e-4, 1e-3$\}$, respectively.

For \LESSR~\footnote{\mbox{\url{https://github.com/twchen/lessr}}},
we initially search $d$, $lr$ and the number of GNN layers $l$ from $\{32, 64, 128\}$, $\{$1e-4, 1e-3$\}$ and $\{1, 2, 3, 4, 5\}$, respectively.

For \DHCN~\footnote{\mbox{\url{https://github.com/xiaxin1998/DHCN}}}, 
we initially search $d$ and the factor for the self-supervision $\beta$ from $\{32, 64, 128\}$ and $\{$1e-4, 1e-3, 1e-2$\}$, respectively.
We use the default number of GNN layers (i.e., 3) 
due to the fact that the original paper of \DHCN shows that \DHCN is not sensitive to this hyper parameter, and it is very expensive to tune hyper parameters for \DHCN (Section~\ref{sec:exp:runtime}).

\ifCLASSOPTIONcaptionsoff
  \newpage
\fi

\bibliographystyle{IEEEtran}
\bibliography{paper}

\begin{IEEEbiographynophoto}{Bo~Peng}
 received his M.S. degree from the Department of Computer and Information Science, 
 Indiana University–Purdue University, Indianapolis, in 2019. He is currently a 
 Ph.D. student at the Computer
 Science and Engineering Department, The Ohio State University.
 His research interests include machine learning, data mining and their applications in
 recommender systems and graph mining.

\end{IEEEbiographynophoto}
\vspace*{-2.5\baselineskip}
\begin{IEEEbiographynophoto}{Chang-Yu Tai}
received his M.S. degree from the Department of Chemistry,
National Taiwan University, in 2018. He is currently an M.S. student at the
Computer Science and Engineering Department, The Ohio State University. His
research interests include deep learning applications in natural language processing
and recommender systems.
\end{IEEEbiographynophoto}
\vspace*{-2.5\baselineskip}
\begin{IEEEbiographynophoto}{Srinivasan~Parthasarathy}
received his Ph.D. degree from the Department of Computer Science,
University of Rochester, Rochester, in 1999. He is currently 
a Professor at the Computer Science and Engineering Department, and 
the Biomedical Informatics Department, The Ohio State University.
His research is on high performance data analytics, graph analytics and network science, and machine learning and database systems.

\end{IEEEbiographynophoto}
\vspace*{-2.5\baselineskip}
\begin{IEEEbiographynophoto}{Xia~Ning}
  received her Ph.D. degree from the Department of Computer Science \& Engineering,
  University of Minnesota, Twin Cities, in 2012. She is currently 
  an Associate Professor at the Biomedical Informatics Department, and the Computer
  Science and Engineering Department, The Ohio State University. Her
  research is on data mining, machine learning and artificial intelligence with applications 
  in recommender systems, drug discovery and medical informatics. 
\end{IEEEbiographynophoto}
\vfill

\end{document}